\documentclass[12pt]{article}
\usepackage{amsmath}
\usepackage{graphicx}
\usepackage{enumerate}
\usepackage{natbib}
\usepackage{url} 

\usepackage{amsmath}
\usepackage{graphicx,psfrag,epsf}
\usepackage{enumerate}
\usepackage{natbib}
\usepackage{url} 
\usepackage{amsthm}
\usepackage{booktabs}
\usepackage{framed}  
\usepackage{caption}
\usepackage{pgfplots}
\usepgfplotslibrary{dateplot}
\usetikzlibrary{snakes}
\usepackage{float}
\usepackage{amsfonts}
\usepackage{multirow, booktabs}
\usepackage{cleveref}
\usepackage{subcaption}
\usepackage[ruled,vlined]{algorithm2e}
\usepackage[T1]{fontenc}
\usepackage[utf8]{inputenc}
\usepackage{authblk}
\usepackage[multiple]{footmisc}
\usepackage{blindtext,titlefoot}
\usepackage{sectsty}
\usepackage{xcolor}
\usepackage{tikz}
\usepackage{amsmath}
\usepackage{graphicx}
\usepackage{comment}
\usepackage{amsfonts}
\usepackage{bbm}
\usepackage{amssymb}
\usepackage{setspace}
\usepackage[figuresright]{rotating}
\usepackage{adjustbox}
\usepackage{pifont}
\usetikzlibrary{positioning, shapes.geometric}

\usepackage{tabularx}
\usepackage{ragged2e} 
\newcolumntype{L}{>{\RaggedRight}X} 
\usepackage{lipsum} 

\usepackage[ruled,vlined]{algorithm2e}
\usepackage{amsmath, amssymb}
\usepackage{amsfonts, multirow, epsfig, subfig}
\usepackage{graphicx, pdflscape, verbatim, enumerate, colortbl, setspace}
\usepackage{setspace, color,bm}
\usepackage[normalem]{ulem}
\usepackage{cite}
\usepackage{multirow}
\usepackage{booktabs,array}
\usepackage{url}
\usepackage{bbm}
\usepackage{bibunits}  

\theoremstyle{definition}

\newtheorem{lemma}{Lemma}

\newtheorem{theorem}{Theorem}

\theoremstyle{definition}
\newtheorem{remark}{Remark}

\newtheorem{assumption}{Assumption}

\makeatletter
\newcommand*{\rom}[1]{\expandafter\@slowromancap\romannumeral #1@}
\makeatother

\def\spacingset#1{\renewcommand{\baselinestretch}%
{#1}\small\normalsize} \spacingset{1}

\AtBeginDocument{%
    \setdisplayskipstretch{1}%

    \everydisplay\expandafter{%
        \the\everydisplay
        \setlength{\abovedisplayskip}{8pt}%
        \setlength{\belowdisplayskip}{6pt}%
        \setlength{\abovedisplayshortskip}{5pt}%
        \setlength{\belowdisplayshortskip}{3pt}%
    }%
}

\begin{document}

\begin{bibunit}[apalike]

\sectionfont{\bfseries\large\sffamily}%
%
\newcommand*\emptycirc[1][1ex]{\tikz\draw (0,0) circle (#1);} 
\newcommand*\halfcirc[1][1ex]{%
  \begin{tikzpicture}
  \draw[fill] (0,0)-- (90:#1) arc (90:270:#1) -- cycle ;
  \draw (0,0) circle (#1);
  \end{tikzpicture}}
\newcommand*\fullcirc[1][1ex]{\tikz\fill (0,0) circle (#1);} 

\subsectionfont{\bfseries\sffamily\normalsize}%
%


\def\spacingset#1{\renewcommand{\baselinestretch}%
{#1}\small\normalsize} \spacingset{1}

\begin{center}
    \Large \bf Reconciling Overt Bias and Hidden Bias in Sensitivity Analysis for Matched Observational Studies
\end{center}

\begin{center}
  \large $\text{Siyu Heng}^{*, \dagger, 1}$, $\text{Yanxin Shen}^{*, 2}$, and $ \text{Pengyun Wang}^{*, 3}$
\end{center}

\begin{center}
   \large \textit{$^{1}$Department of Biostatistics, New York University}
\end{center}

\begin{center}
   \large \textit{$^{2}$Department of Statistics, Pennsylvania State University}
   \end{center}

\begin{center}
   \large \textit{$^{3}$Data Science Institute, The University of Chicago}
\end{center}

\let\thefootnote\relax\footnotetext{$^{*}$Siyu Heng, Yanxin Shen, and Pengyun Wang contributed equally to this work and are listed alphabetically. Yanxin Shen and Pengyun Wang began contributing to this work as student collaborators under the mentorship of Siyu Heng during their master's and undergraduate studies, respectively.}

\begin{abstract}

Matching is one of the most widely used causal inference designs in observational studies, but post-matching confounding bias remains a critical concern. This bias includes \textit{overt bias} from inexact matching on measured confounders and \textit{hidden bias} from unmeasured confounders. Researchers routinely apply the famous Rosenbaum-type sensitivity analysis after matching to assess the impact of these biases on causal conclusions. In this work, we show that this approach is often conservative and may overstate sensitivity to confounding bias because the classical solution to the Rosenbaum sensitivity model may allocate hypothetical hidden bias in ways that contradict the overt bias observed in the matched dataset. To address this problem, we propose a new approach to Rosenbaum-type sensitivity analysis by ensuring compatibility between hidden and overt biases. Our approach does not need to add any additional assumptions (beyond mild regularity conditions) to Rosenbaum-type sensitivity analysis, and can produce uniformly more informative sensitivity analysis results than the conventional Rosenbaum-type sensitivity analysis. Computationally, our approach can be solved efficiently via iterative convex programming. Extensive simulations and a real data application demonstrate substantial gains in statistical power of sensitivity analysis. Importantly, our approach can also be applied to many other sensitivity analysis frameworks.

\end{abstract}

\noindent%
{\it Keywords:} Causal inference; Matching; Randomization test; Randomization-based inference; Rosenbaum bounds.

\spacingset{1.73} 

\section{Introduction}

In observational studies, matching is one of the most important and widely used study designs for causal inference. Ideally, treated and control units are exactly matched on all confounders, rendering treatment assignments as-if randomized within matched sets and enabling valid randomization-based inference, as in a randomized experiment \citep{rosenbaum2002observational, rosenbaum2020design, ding2024first, small2024protocols}. In practice, however, post-matching confounding bias typically exists: post-matching overt bias arises from inexact matching on measured confounders, while post-matching hidden bias results from unmeasured confounders. These biases cause post-matching treatment assignments to deviate from true randomization, undermining the validity of downstream randomization-based inference. To address this, the Rosenbaum-type sensitivity analysis \citep{Rosenbaum1987, rosenbaum2002observational} has become a standard tool for quantifying the impact of confounding bias in matched observational studies. It introduces a uniform sensitivity parameter $\Gamma \geq 1$, which bounds the pairwise ratio of post-matching treatment assignment probabilities within each matched set. The parameter $\Gamma$ reflects the degree of departure from randomization: the larger the $\Gamma$, the greater the potential for post-matching confounding bias. This bounding constraint, known as the \textit{Rosenbaum bounds}, defines the feasible set of post-matching treatment assignment probabilities under confounding bias. Given a specified $\Gamma$, researchers compute the worst-case (maximal) $p$-value or confidence interval over all post-matching treatment assignment probabilities that satisfy the Rosenbaum bounds, or equivalently, over all possible allocations of post-matching confounding bias (i.e., an aggregation of overt and hidden biases not adjusted by matching) (\citealp{Rosenbaum1987, rosenbaum2002observational, rosenbaum2020design}). 

Since its establishment, Rosenbaum-type sensitivity analysis has been widely used in matched observational studies and is highly influential in causal inference more broadly. For representative applications and extensions, see, for example, \citet{Rosenbaum1987, rosenbaum2002observational, rosenbaum2020design}, \citet{heller2009split}, \citet{baiocchi2010building}, \citet{zhang2011using}, \citet{fogarty2016discrete}, \citet{fogarty2019extended}, \citet{karmakar2019integrating}, \citet{zhao2019sensitivityvalue}, \citet{zhang2021selecting}, \citet{ding2024first}, \citet{small2024protocols}, \citet{pimentel2024covariate}, and \citet{wu2025sensitivity}, among many others. Beyond matched observational studies, Rosenbaum-type sensitivity analysis has also been adapted to other study designs, including stratification \citep{rosenbaum2018sensitivity} and instrumental variable settings \citep{kang2016full, ertefaie2018quantitative}.

In this work, we show that unless the matching was exact (which is rare when there are multiple or continuous measured confounders), the conventional Rosenbaum-type sensitivity analysis may be overly conservative, potentially overstating the sensitivity of causal conclusions to post-matching confounding bias. The issue arises because the allocations of hidden bias allowed under the Rosenbaum bounds may contradict overt bias evident in the matched dataset, and we propose to detect such contradiction by using post-matching confounder imbalance information (i.e., post-matching overt bias information) as a valid negative control. To address this critical issue, we introduce a new formulation of the Rosenbaum-type sensitivity analysis that enforces compatibility between overt and hidden biases. Specifically, we replace the original feasible set (defined by the Rosenbaum bounds constraint) with a refined feasible set that incorporates information from the observed post-matching confounder imbalance. Our proposed approach has several remarkable strengths: (i) \textbf{Robust Validity:} The statistical validity of our proposed approach does not depend on any modeling assumptions on treatment assignments. In other words, our approach does not need to add any additional assumptions to the classic Rosenbaum-type sensitivity analysis, except for some mild regularity conditions. (ii) \textbf{Improved Power:} Our proposed approach is uniformly more powerful than the conventional approach to Rosenbaum-type sensitivity analysis, regardless of the underlying data-generating processes. Both the simulation study in Section~\ref{sec: simulations} and a real data application in Section~\ref{sec: application} confirm substantial gains in power using our approach. (iii) \textbf{Efficient Computation:} We show that our proposed approach can be efficiently implemented via an iterative convex programming procedure. A rigorous theoretical justification of computational feasibility is presented in Section~\ref{sec: methods}. (iv) \textbf{Broad Applicability:} Our approach can be applied to general matching designs, including, but not limited to, pair matching, matching with multiple controls, and full matching. Also, our approach is not tied to any specific sensitivity analysis frameworks and can be extended to many other sensitivity analysis settings (see Remark~\ref{rem: extension} in Appendix B).

\section{Review}

\subsection{An Ideal Matched Observational Study Without Post-Matching Biases}\label{subsec: matching}

We follow the classic framework and notations for matched observational studies (\citealp{rosenbaum2002observational, rosenbaum2020design}). Consider a general matched observational study with $I$ matched sets, where there are $n_{i}$ units in matched set $i$ ($i=1,\dots, I$). Let $N=\sum_{i=1}^{I}n_{i}$ denote the total number of units. In each matched set $i$, there are $m_{i}$ treated units and $n_{i}-m_{i}$ control units, where $\min \{m_{i}, n_{i}-m_{i}\}=1$ for all $i$. This general setting covers various matching designs \citep{rosenbaum2002observational, rosenbaum2020design}. For example, if $n_{i}=2$ and $m_{i}=n_{i}-m_{i}=1$ for all $i$, the matching design is pair matching. When $m_{i}=1$ for all $i$ and $n_{i}-m_{i}\geq 2$ for some $i$, the matching design is called matching with multiple (possibly variable) controls. If $\min \left\{{m_{i},n_{i}-m_{i}}\right\} =1$ for all $i$ is the only constraint, the matching design is called full matching. For unit $j$ in matched set $i$, let $Z_{ij}\in \{0,1\}$ denote its treatment indicator ($Z_{ij}=1$ if receiving the treatment and $Z_{ij}=0$ if receiving the control), $\mathbf{x}_{ij}$ its measured confounders, $\mathbf{u}_{ij}$ its unmeasured confounders, and $Y_{ij}$ its observed outcome. Then, we let $\mathbf{Z}=(Z_{11}, \dots, Z_{In_{I}})$ denote the treatment indicator vector, $\mathbf{X}=(\mathbf{x}_{11}, \dots, \mathbf{x}_{In_{I}})$ the collection of all measured confounders, $\mathbf{U}=(\mathbf{u}_{11}, \dots, \mathbf{u}_{In_{I}})$ the collection of all unmeasured confounders, and $\mathbf{Y}=(Y_{11},\dots, Y_{In_{I}})$ the observed outcome vector. Following the potential outcomes framework (\citealp{neyman1923application, rubin1974estimating}), for unit $j$ in matched set $i$, we have $Y_{ij}=Z_{ij}Y_{ij}(1)+(1-Z_{ij})Y_{ij}(0)$, where $Y_{ij}(1)$ denotes its potential outcome under treatment and $Y_{ij}(0)$ its potential outcome under control. Let $\mathcal{Z}=\{\mathbf{Z}\in \{0,1\}^{N}: \sum_{j=1}^{n_{i}}Z_{ij}=m_{i} \text{ for each $i$}\}$ denote the collection of all possible realizations of treatment assignments in the matched dataset.

Assuming exact matching on measured confounders (i.e., $\mathbf{x}_{ij}=\mathbf{x}_{ij^{\prime}}$ for all $i, j, j^{\prime}$) and no unmeasured confounders, the treatments are as-if randomly assigned within each matched set \citep{rosenbaum2002observational, rosenbaum2020design}: 
    \begin{equation}\label{eqn: randomization assumption}
   \text{pr}(Z_{ij}=1 \mid \mathcal{Z}, \mathbf{X})=m_{i}/n_{i}, \text{ for $i=1,\dots, I, j=1,\dots, n_{i}$}.
\end{equation}
Based on the randomization condition (\ref{eqn: randomization assumption}), researchers can use randomization-based inference (as in a randomized experiment) for causal research (\citealp{rosenbaum2002observational, rosenbaum2020design,  ding2024first}). For example, consider the following general family of causal effect mechanisms:
\begin{equation}\label{eqn: general treatment effect model}
   Y_{ij}(1)=f(Y_{ij}(0), \boldsymbol{\beta}, \mathbf{x}_{ij}), \text{ for all } i, j, 
\end{equation}
where $f: Y_{ij}(0)\mapsto Y_{ij}(1)$ can be any invertible function that involves the causal parameter(s) $\boldsymbol{\beta}$ and possibly also the measured confounders $\mathbf{x}_{ij}$. For example, when $f=Y_{ij}(0)+\beta$, model (\ref{eqn: general treatment effect model}) reduces to the constant effect (\citealp{pw1988causal, rosenbaum2002observational, guo2018confidence}). If $f$ includes interaction terms between $\boldsymbol{\beta}$ and $\mathbf{x}_{ij}$, model (\ref{eqn: general treatment effect model}) allows for heterogeneous/individual treatment effects (\citealp{rosenbaum2020design, zhang2021selecting}). Then, we let $H_{\boldsymbol{\beta}_{0}}: \boldsymbol{\beta}=\boldsymbol{\beta}_{0}$ denote a general Fisher's sharp null hypothesis claiming that the causal parameters $\boldsymbol{\beta}$ equal some prespecified value/vector $\boldsymbol{\beta}_{0}$. In the special case of $f=Y_{ij}(0)$, the $H_{\boldsymbol{\beta}_{0}}$ reduces to Fisher's sharp null of no effect $H_{0}: Y_{ij}(1)=Y_{ij}(0) \text{ for all } i, j$ (\citealp{fisher1937design, rosenbaum2002observational}). For a general $H_{\boldsymbol{\beta}_{0}}$, define the adjusted outcome $Y_{ij}^{\text{adj}}=Z_{ij}Y_{ij}+(1-Z_{ij})f(Y_{ij}, \boldsymbol{\beta}_{0}, \mathbf{x}_{ij})$, and let $Y_{ij}^{\text{adj}}(1)$ and $Y_{ij}^{\text{adj}}(0)$ denote the potential outcomes of $Y_{ij}^{\text{adj}}$ under treatment and control, respectively. Under $H_{\boldsymbol{\beta}_{0}}$, we have $Y_{ij}^{\text{adj}}(1)=Y_{ij}^{\text{adj}}(0)=Y_{ij}(1)$, implying that testing $H_{\boldsymbol{\beta}_{0}}$ with $Y_{ij}$ is procedurally equivalent to testing $H_{0}$ with $Y_{ij}^{\text{adj}}$. Also, a valid confidence set for the causal parameters $\boldsymbol{\beta}$ can be obtained via inverting randomization tests for $H_{\boldsymbol{\beta}_{0}}$ under various $\boldsymbol{\beta}_{0}$. In other words, if we can develop a valid test for Fisher's sharp null of no effect $H_{0}$, we immediately obtain a valid test for general Fisher's sharp null hypotheses $H_{\boldsymbol{\beta}_{0}}$, as well as a valid confidence set for causal parameters $\boldsymbol{\beta}$ (\citealp{rosenbaum2002observational, rosenbaum2020design}). Therefore, without loss of generality, we focus on testing $H_{0}$ when illustrating our approach, while noting that all results apply immediately to general causal effects of the form (\ref{eqn: general treatment effect model}).

Then, consider the general family of sum test statistics with the form $T(\mathbf{Z}, \mathbf{Y})=\sum_{i=1}^{I}\sum_{j=1}^{n_{i}}Z_{ij}q_{ij}$, where $q_{ij}$ can be any function (score) based on $\mathbf{Y}$. For example, when $q_{ij}=Y_{ij}$, the test statistic $T$ is the permutational $t$-test if $Y_{ij}$ is continuous or the Mantel-Haenszel test when $Y_{ij}$ is binary. When $q_{ij}=N^{-1}\cdot \text{rank}(Y_{ij})$, the $T$ is the Wilcoxon rank sum test with normalized rank scores. See \citet{rosenbaum2002observational, rosenbaum2020design} for other commonly used sum test statistics such as $U$-statistics and sign-score statistics. Applying the finite-population central limit theorem \citep{rosenbaum2002observational, li2017general}, we have $\{T-E(T)\}/\sqrt{\text{var}(T)} \xrightarrow{d} N(0,1)$ under $H_{0}$, where $E(T)=\sum_{i=1}^{I}\sum_{j=1}^{n_{i}}\frac{m_{i}}{n_{i}}q_{ij}$ and $\text{var}(T)=\sum_{i=1}^{I}\sum_{j=1}^{n_{i}}\frac{1}{n_{i}}q_{ij}^{2}-\sum_{i=1}^{I}(\sum_{j=1}^{n_{i}}\frac{1}{n_{i}}q_{ij})^{2}$, enabling valid randomization-based causal inference \citep{rosenbaum2002observational, rosenbaum2020design}.

\subsection{The Conventional Rosenbaum-Type Sensitivity Analysis}\label{subsec: rosenbaum bounds}

In practice, matching is rarely perfect, and post-matching confounding bias typically remains in matched observational studies. This confounding bias consists of (i) \textit{overt bias} due to inexact matching on measured confounders (especially when there are multiple or continuous measured confounders) and (ii) \textit{hidden bias} due to unmeasured confounders. Together, overt and hidden biases can cause post-matching treatment assignments to depart from random assignments (i.e., the randomization condition (\ref{eqn: randomization assumption}) does not hold) and thereby bias randomization-based inference. To address this concern, researchers commonly adopt the Rosenbaum-type sensitivity analysis framework to assess how sensitive randomization-based inference is to post-matching confounding bias. Specifically, for each unit $j$ in matched set $i$, if $m_{i}=1$ (i.e., matched set $i$ has one treated unit and one or multiple controls), we let $p_{ij}=\text{pr}(Z_{ij}=1 \mid \mathcal{Z}, \mathbf{X}, \mathbf{U})$ represent its probability of receiving the treatment after matching; if $m_{i}>n_{i}-m_{i}=1$ (i.e., matched set $i$ has one control and multiple treated units), we let $p_{ij}=\text{pr}(Z_{ij}=0 \mid \mathcal{Z}, \mathbf{X}, \mathbf{U})$ represent its probability of receiving the control after matching. Henceforth, we call $p_{ij}$ the post-matching treatment (or control) assignment probability if $m_{i}=1$ (or if $m_{i}>n_{i}-m_{i}=1$). In the presence of post-matching overt and hidden biases, the true post-matching treatment assignment probabilities $p_{ij}$ need not equal $1/n_{i}$. The Rosenbaum-type sensitivity analysis imposes the following bounding constraint on the pairwise ratio of $p_{ij}$ within each matched set (i.e., the Rosenbaum bounds), where $\Gamma\geq 1$ is a sensitivity parameter:
\begin{equation}\label{eqn: Rosenbaum bounds}
\Gamma^{-1} \leq p_{ij}/p_{ij^{\prime}}\leq \Gamma \text{ for all $i, j, j^{\prime}$}.
\end{equation}
We let $\Lambda_{\Gamma}$ denote the collection of all possible values of the true post-matching treatment assignment probabilities $\mathbf{p}=(p_{11},\dots, p_{In_{I}})\in [0,1]^{N}$ that satisfy (\ref{eqn: Rosenbaum bounds}), along with the constraint $\sum_{j=1}^{n_{i}}p_{ij}=1$ for all $i$. When $\Gamma=1$, we have $p_{ij}=1/n_{i}$, which corresponds to uniform treatment assignments as in a randomized experiment. As $\Gamma$ deviates from 1, (\ref{eqn: Rosenbaum bounds}) allows a larger magnitude of post-matching confounding bias (an aggregation of overt and hidden biases), and treatment assignments may deviate further from uniformity. In Rosenbaum-type sensitivity analysis, for each prespecified value of $\Gamma$, researchers calculate the worst-case (maximal) $p$-value for causal effects under (\ref{eqn: Rosenbaum bounds}), and the worst-case (largest) confidence set for causal parameters is obtained by inverting these worst-case $p$-values. Through a series of seminal works, explicit solutions for computing worst-case $p$-values under (\ref{eqn: Rosenbaum bounds}) have been developed for various matching designs and causal null hypotheses (\citealp{Rosenbaum1987, rosenbaum2002observational, gastwirth2000asymptotic, fogarty2020studentized, fogarty2023testing, su2024treatment}). Originally, Rosenbaum-type sensitivity analysis was mainly developed for testing Fisher's sharp null hypotheses \citep{Rosenbaum1987, rosenbaum2002observational, gastwirth2000asymptotic}, such as constant treatment effects (including the zero-effect null for all units stated in $H_{0}$) or causal parameters in a fully specified causal effect mechanism (e.g., equation~(\ref{eqn: general treatment effect model})). Recently, important progress has been made in developing valid Rosenbaum-type sensitivity analysis approaches for other fundamental causal null hypotheses. In particular, in addition to Fisher's sharp null, another fundamental class of causal null hypotheses is Neyman's weak null hypotheses $H_{N,\tau_{0}}:\tau=\tau_{0}$, where $\tau = N^{-1}\sum_{i=1}^{I}\sum_{j=1}^{n_{i}}{Y_{ij}(1)-Y_{ij}(0)}$ (i.e., the sample average treatment effect) and $\tau_{0}$ is a prespecified value (e.g., $\tau_{0}=0$). \citet{fogarty2020studentized} developed an explicit and powerful method for conducting Rosenbaum-type sensitivity analysis to test $H_{N,\tau_{0}}$ in pair-matched observational studies, which was later extended by \citet{fogarty2023testing} to general matching designs, and \citet{su2024treatment} further developed a valid Rosenbaum-type sensitivity analysis approach for quantiles of individual treatment effects. In this work, we mainly focus on testing Fisher's sharp null hypotheses, and in Section~\ref{sec: generalization} we discuss opportunities and challenges for extending our approach to other causal null hypotheses (e.g., Neyman's weak null hypotheses) and to other study designs (e.g., stratification).

In matched observational studies, before proceeding to outcome analyses (e.g., sensitivity analyses), researchers routinely conduct balance tests to assess post-matching confounder balance \citep{rosenbaum2020design}. Widely used balance tests include those based on post-matching standardized differences in means for each confounder, as well as randomization-based balance tests that test whether a prespecified $\Gamma\geq 1$ in (\ref{eqn: Rosenbaum bounds}) can be falsified by the observed pattern of post-matching confounder imbalance \citep{gagnon2019classification, branson2021randomization, chen2023testing}; the special case $\Gamma=1$ is equivalent to testing (\ref{eqn: randomization assumption}). Another important class of post-matching balance tests uses negative control outcome information \citep{rosenbaum2023sensitivity}, which can test (\ref{eqn: Rosenbaum bounds}) by detecting imbalance driven by both measured and unmeasured confounders; see Section~\ref{sec: methods} and Remark~\ref{rem: negative control general} in Appendix B for details. Outcome analyses (including sensitivity analyses) are then conducted conditional on the matched dataset passing these balance tests: for example, if a prespecified $\Gamma$ is not rejected, researchers may conduct sensitivity analysis at that value, whereas if it is rejected, they may increase $\Gamma$ until it cannot be rejected \citep{chen2023testing}. Because many matched observational studies involve both balance tests and outcome tests, there are two perspectives on whether one should adjust for multiplicity arising from conducting both types of tests (both discussed in \citet{rosenbaum2023sensitivity}): the first perspective prioritizes rigorous Type-I error control and recommends adjusting for multiplicity when reporting sensitivity analysis results, whereas the second perspective treats balance tests as exploratory and does not adjust for multiplicity when reporting the sensitivity analysis. As will be discussed in Section~\ref{sec: methods}, our proposed approach improves power under either perspective; however, in our theoretical results and simulation studies, we adopt the first perspective and adjust for multiplicity arising from balance and outcome tests to ensure rigorous Type-I error control.

\section{Sensitivity Analysis Informed by Post-Matching Overt Bias}\label{sec: methods}

The popularity of the Rosenbaum-type sensitivity analysis is largely due to (i) its statistical validity without any modeling assumptions on propensity scores and (ii) the interpretability and elegance of a single sensitivity parameter $\Gamma$. However, one critical limitation of the Rosenbaum-type sensitivity analysis is that it discards the overt bias information (i.e., observed post-matching confounder imbalance) during the outcome analysis. This can cause two problems. First, the sensitivity parameter $\Gamma$ in the Rosenbaum bounds may be understated if it cannot even sufficiently account for the post-matching overt bias (let alone the additional hidden bias aggregated over it). This problem is well understood and has been solved by a series of works (\citealp{gagnon2019classification, branson2021randomization, chen2023testing}), which propose to first test $\Gamma=1$ in (\ref{eqn: Rosenbaum bounds}) based on observed post-matching confounder imbalance information; if the balance test rejected $\Gamma=1$, we gradually increase the value of $\Gamma$ until we identify the minimal $\Gamma$ under which the constraint (\ref{eqn: Rosenbaum bounds}) was not falsified by observed confounder imbalance. Such a change point $\Gamma$ is referred to as the \textit{residual sensitivity value} (\citealp{chen2023testing}) (henceforth denoted as $\Gamma_{*}$), which offers a more suitable baseline $\Gamma$ for the Rosenbaum bounds constraint (\ref{eqn: Rosenbaum bounds}) than setting the baseline $\Gamma=1$. 

Second, for any fixed $\Gamma\geq \Gamma_{*}$, ignoring post-matching overt bias information will often render sensitivity analysis overly conservative because the solution to the worst-case $p$-value under the Rosenbaum bounds constraint (\ref{eqn: Rosenbaum bounds}), i.e., the worst-case allocations of hidden biases (or, equivalently, the worst-case allocations of post-matching treatment assignment probabilities $\mathbf{p}$) allowed under the Rosenbaum bounds constraint (\ref{eqn: Rosenbaum bounds}), may contradict the overt bias information observed in the matched dataset. This second problem \textit{was overlooked in the previous literature} and will be pointed out and addressed in this section. Specifically, we propose a new approach to conducting Rosenbaum-type sensitivity analysis by enforcing compatibility between (a) the allocations of hidden bias (when finding the worst-case $p$-value) and (b) the information of overt bias due to inexact matching, and, meanwhile, retain both the merits (i) and (ii) of the conventional Rosenbaum-type sensitivity analysis. Our approach consists of three steps:

\textit{\textbf{Step 1: Constructing a model-free and asymptotically valid confidence set for the true post-matching treatment assignment probabilities $\mathbf{p}$ using the overt bias information.}} Specifically, we consider a general family of sum statistics of the form $S=\sum_{i=1}^{I}\sum_{j=1}^{n_{i}}Z_{ij}s_{ij}$, where each score $s_{ij}$ is a fixed, pre-specified function that is invariant to the treatment assignment vector $\mathbf{Z}$. Although this class of statistics has the same general mathematical form as those used earlier for the outcome analyses, here it is introduced for a distinct purpose, namely, to assess post-matching overt confounder imbalance and thereby constrain the plausible values for the true post-matching treatment assignment probabilities $\mathbf{p}$.

To reflect overt bias after matching, we construct $s_{ij}$ using measured pre-treatment confounder information. For example, if there is only one measured confounder (i.e., $\mathbf{x}_{ij}$ has one dimension), the $s_{ij}$ can be the measured confounder itself. If the measured confounders $\mathbf{x}_{ij}$ are multi-dimensional, we can define $s_{ij}$ to be some summary measure of the confounder information of $\mathbf{x}_{ij}$. For example, we can set $s_{ij}=\widehat{e}_{ij}=\widehat{e}(\mathbf{x}_{ij})$, where $\widehat{e}$ is the estimated propensity score function (either estimated by external historical data or by cross-fitting, which ensures that the score $s_{ij}=\widehat{e}(\mathbf{x}_{ij})$ does not change with $\mathbf{Z}$ in each matched set; see Remark~\ref{rem: cross fitting} in Appendix B). We can also set $s_{ij}=\widehat{p}_{ij}$, where $\widehat{p}_{ij}$ is the estimated value of $p_{ij}$ (fitted via the historical data or cross-fitting to ensure invariance under different $\mathbf{Z}$; see Remarks~\ref{rem: formulas of pij} and \ref{rem: cross fitting} in Appendix B). Another sensible choice is, among many others, to set $s_{ij}$ as the principal component of $\mathbf{x}_{ij}$ (e.g., constructed by principal component analysis).

The following Lemma~\ref{thm: validity of A} gives a valid and model-free confidence set of $\mathbf{p}$ using any balance test statistic $S$, for which the guarantee of the coverage rate only requires some mild regularity assumptions stated below.

\begin{assumption}[Independent Matched Sets]\label{assump: independence assumption}
The post-matching treatment assignments are independent across different matched sets. 

\end{assumption}

\begin{assumption}[The Lindeberg-Feller Conditions for the Balance Test Statistic $S$]\label{assump: CLT for S}
 Suppose that $\max_{i}\max_{j, j^{\prime}}|s_{ij}-s_{ij^{\prime}}|>0$, i.e., the balance test scores $s_{ij}$ are not constant within at least one matched set. For each $i$, we define $\overline{{S}}_{i}=I^{-1/2}\big\{\sum_{j=1}^{n_{i}}Z_{ij}s_{ij}-E_{\mathbf{p}}(\sum_{j=1}^{n_{i}}Z_{ij}s_{ij})\big\}$. Recall that $\mathbf{p}=(p_{11}, \dots, p_{In_{I}})$ denote the underlying true post-matching treatment assignment probabilities. We have: (i) $\lim_{I\rightarrow\infty } \sum_{i=1 }^{I}E_{\mathbf{p}}(\overline{{S}}_{i}^{2})= \sigma_{S}^{2}>0$; (ii) For any $\epsilon>0$, we have $\lim_{I\rightarrow \infty}\sum_{i =1}^{I}  E_{\mathbf{p}}(\overline{{S}}_{i}^{2}\mathbbm{1}\{\overline{{S}}_{i}^{2}>\epsilon\})= 0$. 
\end{assumption}
\begin{remark}\label{rem: bounded s_ij assumption}
Condition~(ii) in Assumption~\ref{assump: CLT for S} holds, for example, if all $n_i$ and $s_{ij}$ are uniformly bounded by some constant (e.g., when $s_{ij}=\widehat{p}_{ij}$).
\end{remark}

Under Assumptions~\ref{assump: independence assumption} and \ref{assump: CLT for S}, invoking the Lindeberg-Feller central limit theorem, we have $\{S-E_{\mathbf{p} }(S)\}/\sqrt{\text{var}_{\mathbf{p}}(S)}\xrightarrow{d} N(0,1)$ as $I\rightarrow \infty$, where $S=\sum_{i=1}^{I}\sum_{j=1}^{n_{i}}Z_{ij}s_{ij}$. Then, we can obtain the following conclusion stated in Lemma~\ref{thm: validity of A}.

\begin{lemma}\label{thm: validity of A}
 Let $\mathcal{I}_{1}=\{i\in \{1,\dots, I\}: m_{i}=1\}$ and $\mathcal{I}_{2}=\{i\in \{1,\dots, I\}: n_{i}-m_{i}=1 \text{ and } m_{i}\geq 2\}$. We define $\Delta_{1-\alpha^{\prime}}=\big\{\mathbf{p}^{\prime}=(p^{\prime}_{11}, \dots, p^{\prime}_{In_{I}}) \in [0,1]^{N}: \sum_{j=1}^{n_{i}}p^{\prime}_{ij}=1 \text{ for all $i$, } \{S-E_{\mathbf{p}^{\prime}}(S)\}^{2}-\chi_{1-\alpha^{\prime}, 1}^{2}\text{var}_{\mathbf{p}^{\prime}}(S)\leq 0 \big\}$, in which $E_{\mathbf{p}^{\prime}}(S)=\sum_{i\in \mathcal{I}_{1}}\sum_{j=1}^{n_{i}}p^{\prime}_{ij}s_{ij}+\sum_{i\in \mathcal{I}_{2}}\sum_{j=1}^{n_{i}}(1-p^{\prime}_{ij})s_{ij}$ is the expectation of $S$ when $\mathbf{p}=\mathbf{p}^{\prime}$, $\text{var}_{\mathbf{p}^{\prime}}(S)=\sum_{i=1}^{I}\big\{ \sum_{j=1}^{n_i}p^{\prime}_{ij}s_{ij}^2-(\sum_{j=1}^{n_i}p^{\prime}_{ij}s_{ij})^2\big \}$ is the variance of $S$ when $\mathbf{p}=\mathbf{p}^{\prime}$, $\alpha^{\prime}\in (0,\alpha)$ is a prespecified confidence-set noncoverage allocation, and $\chi_{1-\alpha^{\prime}, 1}^{2}$ is the $100(1-\alpha^{\prime})\%$-quantile of $\chi^{2}_{1}$. Under Assumptions~\ref{assump: independence assumption} and \ref{assump: CLT for S}, we have $\lim_{I\rightarrow \infty}\text{pr}(\mathbf{p}\in \Delta_{1-\alpha^{\prime}} \mid \mathcal{Z}, \mathbf{X}, \mathbf{U})=1-\alpha^{\prime}$.
\end{lemma}

That is, $\Delta_{1-\alpha^{\prime}}$ in Lemma~\ref{thm: validity of A} is an asymptotically valid confidence set for the unknown true $\mathbf{p}$, even if the score $s_{ij}$ is constructed from a misspecified model for $p_{ij}$ and even in the presence of unmeasured confounders. This is because the validity of $\Delta_{1-\alpha^{\prime}}$ relies only on (i) $s_{ij}$ being fixed with respect to the post-matching treatment indicators $\mathbf{Z}$ and (ii) mild regularity conditions ensuring that a finite-population central limit theorem holds for $S$.

Some specific families of tests $S$ have been considered in relevant works, but were used for different purposes and/or in different ways. For example, \citet{rosenbaum1992detecting, rosenbaum2023sensitivity} propose to set $s_{ij}$ as some negative control outcome (i.e., some outcome unaffected by the treatment) to construct a confidence set for $\mathbf{p}$. However, in many real-world datasets, such a perfect negative control outcome may not exist or may be hard to identify or validate. Therefore, at a high level, what we propose is to use post-matching confounder imbalance information as a negative control, which is always valid (because it consists of pre-treatment information only and is known to be unaffected by the treatment) and almost always exists in the dataset (unless the matching was exact). Some studies also propose to set score $s_{ij}$ in $S$ as some summary measure of confounder imbalance as discussed above (\citealp{gagnon2019classification, branson2021randomization, chen2023testing}), but they only used $S$ to test the randomization condition (i.e., $\Gamma=1$ in (\ref{eqn: Rosenbaum bounds})) or the biased randomization condition (i.e., some prespecified $\Gamma>1$ in (\ref{eqn: Rosenbaum bounds})). In other words, they only used $S$ to study some summary balance measure of $\mathbf{p}=(p_{11},\dots, p_{In_{I}})$ and did not look at the full information of $\mathbf{p}$ captured by the confidence set $\Delta_{1-\alpha^{\prime}}$. Another relevant work is \citet{pimentel2024covariate}, which also proposed to incorporate overt bias information into the Rosenbaum-type sensitivity analysis. However, the validity of the sensitivity analysis method proposed in \citet{pimentel2024covariate} requires additional knowledge on the propensity score model (see Remark~\ref{rem: difference with other sensitivity analysis in inexact matching}). In contrast, our proposed confidence set $\Delta_{1-\alpha^{\prime}}$ has a robust validity guarantee, even when the propensity score models were misspecified when constructing each $s_{ij}$.

\textit{\textbf{Step 2: Finding the worst-case (maximal) $p$-value over all possible post-matching treatment assignment probabilities within the intersection of the Rosenbaum bounds constraint (3) and the confidence set constructed in Step 1.}} That is, when solving the worst-case $p$-value, we replace the conventional feasible set of $\mathbf{p}$ (i.e., the $\Lambda_{\Gamma}$ induced by the Rosenbaum bounds (\ref{eqn: Rosenbaum bounds})) with a refined feasible set $\Lambda_{\Gamma}\cap \Delta_{1-\alpha^{\prime}}$, which considers both the Rosenbaum bounds (\ref{eqn: Rosenbaum bounds}) prespecified by researchers and the confidence set $\Delta_{1-\alpha^{\prime}}$ constructed from the observed data. Compared with the conventional feasible set $\Lambda_{\Gamma}$, the refined feasible set $\Lambda_{\Gamma}\cap \Delta_{1-\alpha^{\prime}}$ is uniformly smaller (i.e., more informative) without any additional modeling assumptions (since the validity of $\Delta_{1-\alpha^{\prime}}$ is model-free). It enforces the allocations of hidden biases allowed under the Rosenbaum bounds (\ref{eqn: Rosenbaum bounds}) to be compatible with the overt bias information captured by $\Delta_{1-\alpha^{\prime}}$. 

If $\Lambda_{\Gamma}\cap \Delta_{1-\alpha^{\prime}}= \emptyset$ (which can be determined by solving a convex program; see Remark~\ref{rem: judge the empty set} in Appendix B), this suggests that researchers may have understated the sensitivity parameter $\Gamma$ as it cannot even sufficiently cover the overt bias information contained in $\Delta_{1-\alpha^{\prime}}$, let alone the potential hidden bias added over it. In this case, researchers should gradually increase $\Gamma$ until $\Lambda_{\Gamma}\cap \Delta_{1-\alpha^{\prime}}\neq \emptyset$. In particular, there is a change point $\Gamma$ such that $\Lambda_{\Gamma}\cap \Delta_{1-\alpha^{\prime}}$ transitions from an empty set to a non-empty set. Such a change point $\Gamma$ agrees with the concept of residual sensitivity value $\Gamma_{*}$ proposed in \citet{chen2023testing}. However, in the previous literature, after identifying the $\Gamma_{*}$ using some explicit methods, people still use the conventional formulation of the Rosenbaum-type sensitivity analysis and ignore the overt bias information contained in scores $s_{ij}$ during the outcome analysis. In contrast, our new formulation overcomes this limitation.

To formally introduce our problem formulation, by the central limit theorem (normal approximation), we have $\{T-E_{\mathbf{p}}(T)\}/\sqrt{\text{var}_{\mathbf{p}}(T)}\xrightarrow{d} N(0,1)$ under $H_{0}$, where $E_{\mathbf{p}}(T)=\sum_{i\in \mathcal{I}_{1}}\sum_{j=1}^{n_{i}}p_{ij}q_{ij}+\sum_{i\in \mathcal{I}_{2}}\sum_{j=1}^{n_{i}}(1-p_{ij})q_{ij}$ and $\text{var}_{\mathbf{p}}(T) =\sum_{i=1}^{I}\big\{ \sum_{j=1}^{n_i}p_{ij}q_{ij}^2-\big(\sum_{j=1}^{n_i}p_{ij}q_{ij}\big)^2
    \big\}$. Therefore, to find the worst-case (maximal) $p$-value among the refined feasible set $\Lambda_{\Gamma}\cap \Delta_{1-\alpha^{\prime}}$ (i.e., our new formulation of Rosenbaum-type sensitivity analysis) is equivalent to solving the worst-case (minimal) squared deviate $\{T-E_{\mathbf{p}}(T)\}^{2}/\text{var}_{\mathbf{p}}(T)$ among $\Lambda_{\Gamma}\cap \Delta_{1-\alpha^{\prime}}$, which can be expressed as the optimal value $d^{*}$ of the following quadratic fractional program:
\begin{equation}\label{eqn: fractional program}
        \text{minimize}_{\mathbf{p}} \ \{T-E_{\mathbf{p}}(T)\}^{2}/\text{var}_{\mathbf{p}}(T) \quad \text{subject to } \mathbf{p}\in \Lambda_{\Gamma}\cap \Delta_{1-\alpha^{\prime}}.
 \end{equation}    

 In contrast, the conventional formulation of Rosenbaum-type sensitivity analysis is equivalent to the following quadratic fractional program:
\begin{equation}\label{eqn: conventional fractional program}
        \text{minimize}_{\mathbf{p}} \ \{T-E_{\mathbf{p}}(T)\}^{2}/\text{var}_{\mathbf{p}}(T) \quad \text{subject to } \mathbf{p}\in \Lambda_{\Gamma}.
 \end{equation}
 We denote the optimal value (i.e., the minimal squared deviate) of (\ref{eqn: conventional fractional program}) by $d_{*}$ and let $\mathbf{p}_{*}$ denote an optimal solution. Since $\Lambda_{\Gamma}\cap\Delta_{1-\alpha^{\prime}}\subseteq\Lambda_{\Gamma}$, we have $d^{*}\geq d_{*}$. The inequality is strict if no optimal solution to (\ref{eqn: conventional fractional program}) belongs to $\Delta_{1-\alpha^{\prime}}$, that is, if all conventional worst-case allocations of $\mathbf{p}$ are incompatible with the overt bias information captured by $\Delta_{1-\alpha^{\prime}}$.

\textit{\textbf{Step 3: Combining the results from Steps 1 and 2 to obtain the final $p$-value reported by our proposed framework.}} For notational convenience, let $\Phi$ denote the standard normal distribution function and $p_{\Delta}^{*}=2\{1-\Phi(\sqrt{d^{*}})\}$ denote the worst-case normal-approximation-based $p$-value over $\mathbf{p}\in\Lambda_{\Gamma}\cap\Delta_{1-\alpha^{\prime}}$, which is obtained from Step 2. In addition, in Steps 1 and 2, under the Rosenbaum bounds constraint (\ref{eqn: Rosenbaum bounds}) with sensitivity parameter $\Gamma$ (i.e., when $\mathbf{p} \in \Lambda_{\Gamma}$), we have established that $\lim_{I\rightarrow \infty}\text{pr}(\mathbf{p}\in \Lambda_{\Gamma}\cap \Delta_{1-\alpha^{\prime}} \mid \mathcal{Z}, \mathbf{X}, \mathbf{U})=\lim_{I\rightarrow \infty}\text{pr}(\mathbf{p}\in  \Delta_{1-\alpha^{\prime}} \mid \mathcal{Z}, \mathbf{X}, \mathbf{U})=1-\alpha^{\prime}$. To obtain the final $p$-value reported by our framework from these two facts, there are two general perspectives, both discussed in Section 5.2 of \citet{rosenbaum2023sensitivity}. The first perspective emphasizes rigorous Type-I error control. Under this perspective, because our sensitivity analysis procedure involves two hypothesis tests, the balance test $S$ used to construct the confidence set $\Delta_{1-\alpha^{\prime}}$ in Step 1 and the outcome test $T$ used to compute $p^{*}_{\Delta}$ from Step 2, we follow \citet{berger1994p} and apply a Bonferroni adjustment to account for multiplicity. Specifically, the final $p$-value reported by our framework is obtained by summing two terms: (i) the worst-case (maximal) $p$-value over $\mathbf{p}\in \Lambda_{\Gamma}\cap \Delta_{1-\alpha^{\prime}}$, denoted as $p_{\Delta}^{*}=2\{1-\Phi(\sqrt{d^{*}})\}$, by Step 2; and (ii) the probability that $\mathbf{p}\notin \Lambda_{\Gamma}\cap \Delta_{1-\alpha^{\prime}}$, which asymptotically equals $\alpha^{\prime}$ by Step 1. Accordingly, we define $p^{*}=2\{1-\Phi(\sqrt{d^{*}})\}+\alpha^{\prime}$. Moreover, under this perspective, for studies that adopt conventional Rosenbaum-type sensitivity analyses, the same multiplicity issue arises whenever any confounder (covariate) balance assessments or balance tests are conducted after matching (which is typically the case), as the downstream outcome test is implicitly conditional on the matched dataset passing such assessments. In this setting, because $d^{*}\geq d_{*}$, our proposed formulation in~(\ref{eqn: fractional program}) uniformly improves upon the conventional formulation~(\ref{eqn: conventional fractional program}) in the sense that, under the same multiplicity adjustment method and the same allocation level $\alpha^{\prime}$ for balance assessment, the resulting worst-case $p$-value $p^{*}$ based on $d^{*}$ is no larger than the worst-case $p$-value based on $d_{*}$. We adopt this perspective throughout the remainder of the manuscript. The second perspective treats balance assessment or test as exploratory and therefore does not adjust for multiplicity in the outcome analysis (e.g., by directly setting $p^{*}=p^{*}_{\Delta}$); under this perspective as well, because $d^{*}\geq d_{*}$, our formulation~(\ref{eqn: fractional program}) yields uniformly more informative $p$-values than those produced by the conventional formulation~(\ref{eqn: conventional fractional program}).

Theorem~\ref{thm: validity} demonstrates the validity of $p$-value $p^{*}$ reported by our proposed framework, based on an additional regularity condition on the outcome test $T$ stated in Assumption~\ref{assump: CLT for T}:

\begin{assumption}[The Lindeberg-Feller Conditions for the Outcome Test Statistic $T$] \label{assump: CLT for T}
Suppose that $\max_{i}\max_{j, j^{\prime}}|q_{ij}-q_{ij^{\prime}}|>0$, i.e., the outcome test scores $q_{ij}$ are not constant within at least one matched set. Recall that $\mathbf{p}=(p_{11},\dots,p_{In_{I}})$ denote the true post-matching treatment assignment probabilities. For each $i$, let $\overline{{T}}_{i}=I^{-1/2}\big\{\sum_{j=1}^{n_{i}}Z_{ij}q_{ij}-E_{\mathbf{p}}(\sum_{j=1}^{n_{i}}Z_{ij}q_{ij})\big\}$. We have: (i) $\lim_{I\rightarrow\infty} \sum_{i=1 }^{I}E_{\mathbf{p}}(\overline{{T}}_{i}^{2})= \sigma_{T}^{2}>0$; (ii) For any $\epsilon>0$, we have $\lim_{I\rightarrow \infty}\sum_{i=1 }^{I}E_{\mathbf{p}}(\overline{{T}}_{i}^{2}\mathbbm{1}\{\overline{{T}}_{i}^{2}>\epsilon\})= 0$.
\end{assumption}

Under Assumptions~\ref{assump: independence assumption} and \ref{assump: CLT for T}, invoking the Lindeberg-Feller central limit theorem, under $H_{0}$, we have $ \{T-E_{\mathbf{p} }(T)\}/\sqrt{\text{var}_{\mathbf{p}}(T)}\xrightarrow{d} N(0,1)$ as $I\rightarrow\infty$.

\begin{theorem}\label{thm: validity}
Under Assumptions \ref{assump: independence assumption}--\ref{assump: CLT for T}, for any $\alpha\in (0,1)$, for any $\mathbf{p}$ that satisfy the Rosenbaum bounds constraint (\ref{eqn: Rosenbaum bounds}), we have $\limsup_{I\rightarrow \infty} \text{pr}_{\mathbf{p}}(p^{*}\leq \alpha \mid H_{0})\leq \alpha$.

\end{theorem}

Having discussed the validity and efficiency gains of our new sensitivity analysis formulation, the final remaining question is: how to solve the new formulation expressed as the nonlinear fractional program (\ref{eqn: fractional program})? Unlike the conventional formulation (\ref{eqn: conventional fractional program}), the new formulation (\ref{eqn: fractional program}) does not have an explicit solution. Fortunately, we will show that the nonlinear fractional program (\ref{eqn: fractional program}) can be efficiently solved by an iterative convex programming approach, of which the formal procedure is described in Algorithm~\ref{alg: part with covariate adjustment} below. Specifically, given any $c\geq 0$, we consider:
\begin{equation}\label{eqn: convex program}
        \text{minimize}_{\mathbf{p}} \ \{T-E_{\mathbf{p}}(T)\}^{2}-c\times \text{var}_{\mathbf{p}}(T) \quad \text{subject to } \mathbf{p}\in \Lambda_{\Gamma}\cap \Delta_{1-\alpha^{\prime}},
 \end{equation}
of which the explicit form can be expressed as the following quadratic program:
\begin{equation*}
     \begin{split}
        \underset{\mathbf{p}}{\text{minimize}} \quad & \{T-E_{\mathbf{p}}(T)\}^{2}-c\times \text{var}_{\mathbf{p}}(T) \quad \quad (L_{c}) \\
         \text{subject to}\quad &\sum_{j=1}^{n_{i}} p_{ij}=1, \quad\forall i,\\
        &0 \leq p_{ij}\leq 1, \quad \forall i, j,\\
        & p_{ij}-\Gamma p_{ij^{\prime}}\leq 0, \quad \forall i, j, j^{\prime},\\
        & \{S-E_{\mathbf{p}}(S)\}^{2}-\chi_{1-\alpha^{\prime}, 1}^{2}\text{var}_{\mathbf{p}}(S) \leq 0.
     \end{split}
 \end{equation*}
By generalizing the arguments adopted in \citet{rosenbaum1992detecting, rosenbaum2023sensitivity}, we can prove the convexity of the quadratic program (\ref{eqn: convex program}):
 \begin{theorem}\label{thm: convexity}
     For any fixed $c\geq 0$, the quadratic program (\ref{eqn: convex program}) is convex.  
\end{theorem}


Then, Algorithm~\ref{alg: part with covariate adjustment} presents an iterative convex programming approach for solving the proposed formulation~(\ref{eqn: fractional program}) of Rosenbaum-type sensitivity analysis. The primary difficulty is that the objective function in~(\ref{eqn: fractional program}) is a quadratic fractional function, which is generally challenging to optimize directly. Building on the classic Dinkelbach method for nonlinear fractional programming \citep{dinkelbach1967nonlinear}, Algorithm~\ref{alg: part with covariate adjustment} reformulates the original quadratic fractional program into a sequence of parametric quadratic programs of the form~(\ref{eqn: convex program}), indexed by a scalar parameter $c$. At each iteration, $c$ is fixed at its current value, yielding a tractable convex quadratic program (with convexity established in Theorem~\ref{thm: convexity}), whose solution is then used to update $c$. This iterative procedure guarantees monotone improvement toward the optimum of the original fractional program (\ref{eqn: fractional program}), denoted by $d^{*}$, and continues until convergence (the detailed proof can be found in Appendix A). Combined with Theorem~\ref{thm: validity}, Algorithm~\ref{alg: part with covariate adjustment} therefore provides an efficient and trackable method for computing the worst-case $p$-value under the proposed sensitivity analysis formulation~(\ref{eqn: fractional program}).

\begin{algorithm}  
\SetAlgoLined

\caption{An iterative convex programming approach to Rosenbaum-type sensitivity analysis informed by post-matching overt bias information.} \label{alg: part with covariate adjustment}

\KwIn{The observed data $(\mathbf{Z}, \mathbf{X}, \mathbf{Y})$, the sensitivity parameter $\Gamma$ in the Rosenbaum bounds (\ref{eqn: Rosenbaum bounds}), and some prespecified $\alpha^{\prime}\in (0, \alpha)$ adopted in $\Delta_{1-\alpha^{\prime}}$ (e.g., set $\alpha^{\prime}=\alpha/2$, where $\alpha$ is the significance level). }

\smallskip
\smallskip
\begin{enumerate}
  \setcounter{enumi}{-1}

  \item If $\Lambda_{\Gamma}\cap \Delta_{1-\alpha^{\prime}}\neq \emptyset$, we proceed to the next step. Otherwise, we increase $\Gamma$ until $\Lambda_{\Gamma}\cap \Delta_{1-\alpha^{\prime}}\neq \emptyset$.

  \item  We set an initial value $c_{(0)}>0$ (e.g., set $c_{(0)}=\chi^{2}_{1-\alpha^{\prime}, 1}$).

\item In each iteration $m \geq 1$, find the optimal solution $\mathbf{p}_{(m)}$ to the convex program (\ref{eqn: convex program}) with $c=c_{(m-1)}$. Then, we set $c_{(m)}=\{T-E_{\mathbf{p}_{(m)}}(T)\}^{2}/\text{var}_{\mathbf{p}_{(m)}}(T)$.
\item Repeat iterations until $|c_{(m)}-c_{(m-1)}|<\epsilon$, in which $\epsilon$ is a prespecified small value.  
\end{enumerate}
\textbf{Output:} The worst-case $p$-value $p_{(m)}=2\{1-\Phi(\sqrt{c_{(m)}})\}+\alpha^{\prime}$.
\end{algorithm}

By standard results for nonlinear fractional programming \citep{dinkelbach1967nonlinear}, the iterative update of the parameter $c$ in Algorithm~\ref{alg: part with covariate adjustment} converges to the optimal value $d^{*}$ of the original fractional program (\ref{eqn: fractional program}). The following theorem formalizes this convergence result.
\begin{theorem}\label{thm: convergence}
Under Assumptions~\ref{assump: independence assumption}--\ref{assump: CLT for T}, for Algorithm~\ref{alg: part with covariate adjustment}, we have $c_{(m)}\rightarrow d^{*}$ and 
$p_{(m)}\rightarrow p^{*}=2\{1-\Phi(\sqrt{d^{*}})\}+\alpha^{\prime}$ as the number of iterations $m\rightarrow \infty$.
\end{theorem}

In summary, the proposed approach is asymptotically valid (proved in Theorem~\ref{thm: validity}), uniformly more informative than the conventional approach (because $d^{*}\geq d_{*}$) without adding any assumptions to the Rosenbaum-type sensitivity analysis framework (except for some mild regularity conditions to ensure the asymptotic normality of the balance test $S$), and computationally feasible (proved in Theorems~\ref{thm: convexity} and \ref{thm: convergence}). Also, note that the idea of our approach (i.e., using post-matching overt bias information as a valid negative control to refine the feasible set of hidden biases in a sensitivity analysis) is not tied to any specific sensitivity analysis frameworks (because $\Delta_{1-\alpha^{\prime}}$ does not depend on any sensitivity models). In Remark~\ref{rem: extension} in Appendix B, we briefly discuss how our approach can be directly applied to other sensitivity analysis frameworks in matching.

\begin{remark}\label{rem: number of iterations}
 Theorem~\ref{thm: convergence} states that the iterative convex programming procedure described in Algorithm~\ref{alg: part with covariate adjustment} converges to the target value (i.e., the $p$-value $p^{*}$) for some sufficiently large number of iterations. In practice, we find that convergence is achieved rapidly. For example, in the real data application presented in Section~\ref{sec: application}, Algorithm~\ref{alg: part with covariate adjustment} typically converges within a short running time, e.g., within 30 seconds for the considered values of sensitivity parameter $\Gamma$ (see Appendix C in the supplementary materials for details).
 
\end{remark}

\begin{remark}[Differences with Existing Sensitivity Analysis Approaches in Inexactly Matched Observational Studies]\label{rem: difference with other sensitivity analysis in inexact matching}

There are two recent works on sensitivity analysis for confounding biases in inexactly matched observational studies (i.e., \citealp{pimentel2024covariate, li2025sensitivity}). Together with our proposed approach, these three sensitivity analysis methods adopt different setups and target different sources of confounding bias, leading to distinct interpretations of sensitivity analysis results.

Specifically, \citet{pimentel2024covariate} considers the following Rosenbaum-type biased propensity score model \citep{rosenbaum2002observational}: $\text{logit}\ \text{pr}(Z_{ij}=1\mid \mathbf{x}_{ij}, u_{ij})=\kappa(\mathbf{x}_{ij})+\gamma u_{ij}$, where $\text{pr}(Z_{ij}=1\mid \mathbf{x}_{ij}, u_{ij})$ denotes the propensity score (i.e., the population-level, pre-matching treatment assignment probability), $\kappa(\mathbf{x}_{ij})$ is a function of measured confounders $\mathbf{x}_{ij}$, $u_{ij}\in[0,1]$ is an unmeasured confounder normalized to $[0,1]$, and $\gamma\geq 0$ is a prespecified sensitivity parameter that quantifies the magnitude of hidden bias due to $u_{ij}$. When matching is exact (i.e., $\mathbf{x}_{ij}=\mathbf{x}_{ij^{\prime}}$ for all $i,j,j^{\prime}$), all post-matching confounding bias arises from unmeasured confounding through $u_{ij}$, and there is no post-matching overt bias. In this case, \citet{rosenbaum2002observational} showed that the sensitivity parameter $\Gamma$ in the Rosenbaum bounds constraint~(\ref{eqn: Rosenbaum bounds}) satisfies $\Gamma=\exp(\gamma)$. When matching is inexact, \citet{pimentel2024covariate} proposes a sensitivity analysis method for post-matching hidden bias (indexed by $\gamma$), under the assumptions that (i) there is no interaction between $\mathbf{x}_{ij}$ and $u_{ij}$ \citep{heng2021sharpening} and (ii) a bound or estimate of $\kappa(\mathbf{x}_{ij})$ is available. Under these assumptions, post-matching overt bias induced by inexact matching is treated as known or controlled, so the sensitivity analysis focuses exclusively on post-matching hidden bias. Consequently, robustness to a given value of $\gamma$ in \citet{pimentel2024covariate} should be interpreted as robustness to unmeasured confounding conditional on post-matching overt bias being adequately characterized by the additional information of $\kappa(\mathbf{x}_{ij})$ and the absence of interactions between measured and unmeasured confounders, rather than robustness to arbitrary post-matching departures from randomization. In contrast, our proposed approach evaluates robustness to the aggregate effect of post-matching overt and hidden biases, without introducing additional assumptions or external information to characterize overt bias.

By comparison, \citet{li2025sensitivity} proposes a new sensitivity analysis framework for matched observational studies that accommodates both post-matching overt and hidden biases and permits flexible matching designs. Unlike Rosenbaum-type sensitivity analysis models, such as the Rosenbaum bounds~(\ref{eqn: Rosenbaum bounds}) and the biased propensity score models described above, which characterize how post-matching treatment assignment probabilities are distorted by confounding bias, the framework of \citet{li2025sensitivity} models how post-matching randomization-based (permutation-based) distributions of potential outcomes within each matched set are altered by post-matching confounding biases. Because our proposed approach is built on the Rosenbaum bounds sensitivity analysis framework, sensitivity parameters in our framework and in \citet{li2025sensitivity} index different types of post-matching departures from randomization, so robustness conclusions under the two approaches are not directly comparable and reflect protection against distinct forms of post-matching confounding bias. We refer readers to \citet{li2025sensitivity} for a detailed discussion of the connections and distinctions between treatment-assignment-based and potential-outcome-based sensitivity analysis models in matched observational studies.

\end{remark}

\begin{remark}[Connections and Differences with the Data-Compatible Sensitivity Analysis Approaches in the Weighting Literature]\label{rem: differences with other data-compatible sensitivity analysis}

Recent work in the weighting literature, including survey sampling and weighted observational studies, has proposed incorporating auxiliary sample information, such as overt confounder (covariate) imbalance between comparison groups, to obtain more informative partial identification or sensitivity analysis results by restricting the feasible set to be compatible with observed sample characteristics (e.g., \citealp{dorn2023sharp, tudball2023sample}). These approaches are often referred to as data-compatible or sample-constrained sensitivity analyses. In survey sampling, \citet{tudball2023sample} incorporates moment constraints derived from measured confounders into sensitivity analysis for hidden selection bias, while in weighted observational studies, \citet{dorn2023sharp} introduces confounder quantile balancing constraints for sensitivity analysis of inverse probability weighting, together with efficient convex programming formulations and sharpness results under those constraints.

Our work is closely related to, yet distinct from, \citet{dorn2023sharp} and \citet{tudball2023sample}. At a high level, all three approaches share the common principle of leveraging measured confounder balance information to rule out implausible allocations of hidden bias and thereby sharpen sensitivity analysis conclusions, operationalizing this idea by appending balance-induced constraints to the underlying optimization problem. However, there are several essential differences. First, the inferential targets differ: \citet{dorn2023sharp} use the confounder balance information to restrict feasible sets of population-level treatment assignment probabilities (and their sample counterparts), and \citet{tudball2023sample} use such information to constrain population-level coefficients in parametric selection models, whereas our approach uses the confounder balance information to restrict the feasible set of \emph{finite-population, post-matching} treatment assignment probabilities, aligning naturally with a design-based perspective for matched observational studies. Second, existing data-compatible sensitivity analysis approaches for weighting impose balance through quantile constraints \citep{dorn2023sharp} or confounder moment constraints \citep{tudball2023sample}, while we instead use a scalar summary measure $s_{ij}$ of \textit{post-matching} confounder imbalance (e.g., $s_{ij}=\widehat{p}_{ij}$, the post-matching treatment assignment probability estimated from measured confounders). Third, although the resulting optimization problems in \citet{dorn2023sharp} and \citet{tudball2023sample} are convex, our sensitivity analysis approach leads to a non-convex quadratic fractional program (\ref{eqn: fractional program}); nevertheless, we show that it can be solved via a carefully constructed iterative convex programming procedure (Algorithm~\ref{alg: part with covariate adjustment}).

\end{remark}

\section{Simulation Studies}\label{sec: simulations}

We conduct simulation studies to investigate gains in power using the proposed sensitivity analysis approach, compared with the conventional approach. We set the sample size $N=500$ and $N=1000$. The five independent measured confounders $\mathbf{x}_{n}=(x_{n1},\dots, x_{n5})$ and one unmeasured confounder $u_{n}$ for each unit $n$ are generated from the following process: $(x_{n1}, x_{n2}, x_{n3}) \overset{\text{i.i.d.}}{\sim} \mathcal{N}((0,0,0),\mathbf{I}_{3 \times 3})$, $x_{n4}\overset{\text{i.i.d.}}{\sim} \text{Laplace}(0, \sqrt{2}/2)$, $x_{n5}\overset{\text{i.i.d.}}{\sim} \text{Laplace}(0, \sqrt{2}/2)$, and $u_{n}\overset{\text{i.i.d.}}{\sim} N(0, 1)$. Let $\phi(\mathbf{x}_{n})=0.3 (x_{n1} +x_{n2}) +0.6 \cos x_{n3} +0.4 (|x_{n4}|+x_{n5})+|x_{n1}  x_{n2}| +x_{n1} x_{n3}+x_{n4} x_{n5}$. We consider the following three models for the treatment indicator $Z_{n}$: (i) Model 1 (nonlinear logistic model): $\text{logit} \ \text{pr}(Z_{n}=1\mid \mathbf{x}_{n}, u_{n})=\phi(\mathbf{x}_{n})+\gamma u_{n} -2 $; (ii) Model 2 (nonlinear selection model with normal errors): $Z_{n}=\mathbbm{1}\{\phi(\mathbf{x}_{n})+\gamma u_{n} -2>\epsilon_{n1}\}$, where $\epsilon_{n1} \overset{\text{i.i.d.}}{\sim} N(0,1)$; and (iii) Model 3 (nonlinear selection model with non-normal errors): $Z_{n}=\mathbbm{1}\{\phi(\mathbf{x}_{n})+\gamma u_{n} -2>\epsilon_{n1}\}$, where $\epsilon_{n1} \overset{\text{i.i.d.}}{\sim} t_{3}$. We generate $Y_{n}(0)=x_{n1}+|x_{n2}|+0.2x_{n3}^{3}+0.2x_{n5}+ \gamma u_{n} +\epsilon_{n2}$, where $\epsilon_{n2} \overset{\text{i.i.d.}}{\sim} N(0,1)$, and set $Y_{n}(1)=Y_{n}(0)+\beta^*$, where the effect size $\beta^*$ equals $\sqrt{\text{var}(Y_{n}(0))}$. We consider two scenarios of unmeasured confounding: (i) \textit{Scenario 1:} The ``favorable situation'' considered in \citet{rosenbaum2004design, rosenbaum2020design}, in which we set $\gamma=0$. In this scenario, all the post-matching confounding bias is contributed by inexact matching on measured confounders (overt bias). However, researchers would not know this and would still conduct a sensitivity analysis for the hypothetical combined effects of post-matching overt and hidden biases, and would prefer a sensitivity analysis approach with larger power. (ii) \textit{Scenario 2:} The ``unfavorable situation'' in which we set $\gamma=0.5$, representing that, in addition to overt bias due to inexact matching, there exists a large magnitude of hidden bias due to unmeasured confounding. 

After generating $N$ unmatched units in each simulation run, we generate the corresponding matched dataset using the optimal full matching design \citep{hansen2004full, hansen2006optimal}. We only keep the matched datasets that satisfy the common balance criteria that the absolute standardized difference of each measured confounder between the treatment and control groups is below 0.2 \citep{rosenbaum2020design, pimentel2024covariate}. Table~\ref{tab: std} in Appendix B indicates that matching greatly reduced overt bias compared with the unmatched datasets: the average absolute standardized differences in means of all five measured confounders among the 1000 matched datasets are much less than the commonly used threshold of 0.2 (over half of them are less than 0.1). However, there is still an evident residual confounder imbalance after matching (i.e., post-matching overt bias). The conventional Rosenbaum-type sensitivity analysis typically ignores such overt bias information during the outcome analysis stage, as long as the matched dataset passes some balance assessment or balance test \citep{rosenbaum2002observational, rosenbaum2020design}. In contrast, our new formulation incorporates such information into the sensitivity analysis for causal effects.

\begin{table}[ht]
\centering
\footnotesize
\caption{Simulated power of the two approaches to Rosenbaum-type sensitivity analysis: the conventional approach and the proposed new approach. }
\resizebox{\textwidth}{!}{%
\begin{tabular}{cccccccc}
\toprule
& \multirow{2}{*}{Sample Size $N=500$}&\multicolumn{2}{c}{Model 1} &\multicolumn{2}{c}{Model 2} &\multicolumn{2}{c}{Model 3} \\ 
\cmidrule(rl){3-4} \cmidrule(rl){5-6} \cmidrule(rl){7-8}
 & & Scenario 1 & Scenario 2 &  Scenario 1 & Scenario 2 & Scenario 1 & Scenario 2 \\
\midrule
\multirow{3}{*}{$\Gamma=7$} & Conventional & 0.453 & 0.643 & 0.578 & 0.857 & 0.499 & 0.774\\
& Proposed (XGBoost) & 0.497 & 0.691 & 0.821 & 0.938 & 0.634 & 0.846\\
& Proposed (Random Forest) & 0.516 & 0.716 & 0.821 & 0.945 & 0.649 & 0.852\\
\midrule
\multirow{3}{*}{$\Gamma=8$} & Conventional & 0.213 & 0.361 & 0.320 & 0.671 & 0.264 & 0.516\\
& Proposed (XGBoost) & 0.239 & 0.397 & 0.578 & 0.813 & 0.346 & 0.590\\
& Proposed (Random Forest) & 0.249 & 0.408 & 0.562 & 0.812 & 0.375 & 0.624\\
\midrule
& \multirow{2}{*}{Sample Size $N=1000$}&\multicolumn{2}{c}{Model 1} &\multicolumn{2}{c}{Model 2} &\multicolumn{2}{c}{Model 3} \\ 
\cmidrule(rl){3-4} \cmidrule(rl){5-6} \cmidrule(rl){7-8}
 & & Scenario 1 & Scenario 2 &  Scenario 1 & Scenario 2 & Scenario 1 & Scenario 2 \\
\midrule
\multirow{3}{*}{$\Gamma=9$} & Conventional & 0.594 & 0.842 & 0.794 & 0.975 & 0.666 & 0.928\\
& Proposed (XGBoost) & 0.708 & 0.904 & 0.985 & 0.998 & 0.882 & 0.982\\
& Proposed (Random Forest) & 0.770 & 0.932 & 0.991 & 1.000 & 0.929 & 0.990\\
\midrule
\multirow{3}{*}{$\Gamma=10$} & Conventional & 0.364 & 0.647 & 0.553 & 0.907 & 0.436 & 0.799\\
& Proposed (XGBoost) & 0.458 & 0.725 & 0.928 & 0.989 & 0.688 & 0.909\\
& Proposed (Random Forest) & 0.525 & 0.775 & 0.939 & 0.995 & 0.734 & 0.940\\
\midrule 
\multirow{3}{*}{$\Gamma=11$} & Conventional & 0.182 & 0.408 & 0.335 & 0.752 & 0.254 & 0.591\\
& Proposed (XGBoost) & 0.242 & 0.507 & 0.785 & 0.954 & 0.459 & 0.767\\
& Proposed (Random Forest) & 0.288 & 0.559 & 0.822 & 0.967 & 0.506 & 0.814\\
\bottomrule
\end{tabular}}
\label{tab: more simulated power}

\end{table}

We use the permutational $t$-test statistic (i.e., set $q_{ij}=Y_{ij}$) and set $s_{ij}=\widehat{p}_{ij}$, where $\widehat{p}_{ij}$ is estimated using XGBoost or random forest with cross-fitting (see Remark~\ref{rem: cross fitting} in Appendix~B for implementation details). As noted in Section~\ref{sec: methods}, cross-fitting is needed to ensure that $s_{ij}=\widehat{p}_{ij}$ is held fixed with respect to the post-matching treatment assignments $\mathbf{Z}$ in the testing dataset. In addition, cross-fitting can mitigate overfitting when using flexible propensity score models such as XGBoost or random forest. We consider the two approaches to Rosenbaum-type sensitivity analysis: the conventional approach and our proposed new approach described in Algorithm~\ref{alg: part with covariate adjustment}. To ensure that we consider only plausible values of the sensitivity parameter $\Gamma$, prior to applying either the conventional or the proposed approach, we conduct balance tests at significance level $\alpha^{\prime}=\alpha/2$ to identify the residual sensitivity value $\Gamma_{*}$ associated with the Rosenbaum bounds constraint~(\ref{eqn: Rosenbaum bounds}), and restrict attention to sensitivity parameters satisfying $\Gamma\geq\Gamma_{*}$. For fair comparison, both approaches adopt the same Bonferroni-based multiplicity adjustment proposed by \citet{berger1994p} to account for the two tests involved in sensitivity analysis (i.e., the balance test and the outcome test), as discussed in Section~\ref{sec: methods}. Recall that, under this adjustment, the $p$-value reported by our proposed approach is $p^{*}=2\{1-\Phi(\sqrt{d^{*}})\}+\alpha^{\prime}$, while that reported by the conventional approach is $p_{*}=2\{1-\Phi(\sqrt{d_{*}})\}+\alpha^{\prime}$.

 Simulated power in each setting is reported in Table~\ref{tab: more simulated power}, with the simulated Type-I error rate less than the nominal level $\alpha= 0.05$ for both of the two compared approaches across all the considered settings. Table~\ref{tab: more simulated power} shows that our proposed approach can substantially improve power of sensitivity analysis by incorporating the overt bias information in $\Delta_{1-\alpha^{\prime}}$.

\section{Data Application}\label{sec: application}

We use the dataset in \citet{Heller2010UsingTC} to re-analyze the impact of attending a two-year community college (the treatment group) versus a four-year college (the control group) on educational attainment (measured by the total number of years of education). The unmatched dataset contains 1819 students with demographic and socioeconomic information. In particular, the baseline test scores of these students are no less than 55, which is the median test score of students who attended a four-year college. A student who attended a two-year community college and scored at least 55 on the test might plausibly have been admitted to a four-year college, making it meaningful to consider the counterfactual outcome had the student attended a four-year college instead (\citealp{Heller2010UsingTC}). In the unmatched dataset, there is a substantial imbalance in measured confounders between the treatment group (students who attended two-year community colleges) and the control group (students who attended four-year colleges). We perform matching using all 20 measured confounders considered in \citet{Heller2010UsingTC}, such as gender, race, Hispanic ethnicity, baseline test scores, family income, home ownership, the percentage of white students at the respondent's high school, and whether the respondent lived in an urban or rural area, among others. We apply the optimal full matching algorithm with a propensity score caliper (\citealp{hansen2004full}), implemented via the \texttt{R} package \texttt{optmatch} (\citealp{hansen2006optimal}). We limit each matched set to have a maximum of 5 treated units or 5 control units. After optimal full matching, a total of 1819 individuals were optimally grouped into 422 matched sets based on the minimized total rank-based Mahalanobis distance in measured confounders. For all 20 measured confounders, the post-matching absolute standardized differences in means between the treatment and control groups are no greater than 0.2 (most are below 0.1); see Table~\ref{tab: realstd} in Appendix~B for details.

\begin{table}[ht]
\centering
\caption{Sensitivity analysis results for the conventional and proposed approaches.}
\footnotesize
\begin{tabular}{ccccc}
\toprule
\multirow{2}{*}{}&\multicolumn{2}{c}{Conventional} &\multicolumn{2}{c}{Proposed} \\ 
\cmidrule(rl){2-3} \cmidrule(rl){4-5} 
 & $p$-value & 95\% CI & $p$-value & 95\% CI \\
\midrule
$\Gamma=3.50$  & $0.0535$ & $[-2.453, +0.007]$ & $0.0254$ & $[-2.302, -0.191]$ \\
$\Gamma=3.75$  & $0.1000$ & $[-2.503, +0.067]$ & $0.0274$ & $[-2.362, -0.120]$ \\
$\Gamma=4.00$  & $0.1908$ & $[-2.550, +0.123]$ & $0.0343$ & $[-2.418, -0.054]$ \\
$\Gamma=4.25$  & $0.3420$ & $[-2.594, +0.176]$ & $0.0530$ & $[-2.471, +0.006]$ \\
\hline
Sensitivity Value  &  \multicolumn{2}{c}{3.47}   &   \multicolumn{2}{c}{4.22} \\
\hline
\end{tabular}
\label{tab: data analysis}
\end{table}

We applied two approaches to Rosenbaum-type sensitivity analysis: the conventional approach and our proposed approach (Algorithm~\ref{alg: part with covariate adjustment}). For both approaches, we follow the same workflow as in the simulation studies in Section~\ref{sec: simulations}, including conducting balance tests at significance level $\alpha^{\prime}=\alpha/2$ to identify the residual sensitivity parameter and applying the same Bonferroni-based multiplicity adjustment to account for the balance and outcome tests. As in the simulation studies, we set $s_{ij}=\widehat{p}_{ij}$, where $\widehat{p}_{ij}$ is estimated using random forest with cross-fitting (see Remark~\ref{rem: cross fitting} in Appendix B for implementation details). Table~\ref{tab: data analysis} shows that our proposed approach can produce much more informative results (i.e., much smaller $p$-values and confidence intervals) than those reported by the conventional approach. Specifically, the sensitivity value (i.e., the largest $\Gamma$ under which the worst-case $p$-value is below $\alpha=0.05$; see \citealp{zhao2019sensitivityvalue}) under the proposed approach is much larger than that under the conventional approach (4.22 versus 3.47).

\section{Generalizations of the Proposed Approach: Opportunities and Challenges}\label{sec: generalization}

In this section, we outline potential directions for extending the proposed approach to other study designs, such as stratified observational studies, and to alternative causal null hypotheses, including Neyman's weak null hypotheses. We discuss both the opportunities offered by these extensions and the computational challenges that arise in these extensions.

\subsection{Generalizations to Stratified Observational Studies}

In addition to matched observational studies, another important class of observational designs is stratification, in which confounders (covariates) are typically coarsened into categorical values and study units are partitioned into non-overlapping strata based on these coarsened confounders \citep{rosenbaum2002observational, rosenbaum2018sensitivity}. While stratification can effectively control for confounding at the level of the coarsened confounders, it may discard substantial information about post-stratification overt bias, as measured by imbalance in the original, uncoarsened confounders, if such information is not explicitly incorporated into the sensitivity analysis. In this subsection, we discuss both the opportunities and the computational challenges that arise when extending the proposed sensitivity analysis framework from matched to stratified observational studies.

Specifically, consider a stratified observational study with $I$ independent strata. In stratum $i\in \{1,\dots, I\}$, we have $m_{i}$ treated units and $n_{i}-m_{i}$ controls, where $\min\{m_{i}, n_{i}-m_{i}\}\geq 1$. Let $\mathbf{Z}_{i}=(Z_{i1},\cdots, Z_{in_{i}})\in \{0,1\}^{n_{i}}$ denote the treatment indicators for stratum $i$, and $\mathcal{Z}_{i}=\{\mathbf{z}_{i}=(z_{i1},\dots, z_{in_{i}})\in \{0,1\}^{n_{i}}: \sum_{j=1}^{n_{i}}z_{ij}=m_{i}\}$ denote the collection of all possible vector values that $\mathbf{Z}_{i}$ can take, where we have $|\mathcal{Z}_{i}|={n_{i}\choose m_{i}}$. For notational simplicity, we define $l_{i}={n_{i}\choose m_{i}}$ for each $i=1,\dots, I$, and define $L=\sum_{i=1}^{I}l_{i}$. For any $\mathbf{b}_{ik}=(b_{ik1},\dots, b_{ikn_{i}})\in \mathcal{Z}_{i}$, where $i=1,\dots, I$ and $k=1,\dots, l_{i}$, we define $p_{ik}=\text{pr}(\mathbf{Z}_{i}=\mathbf{b}_{ik}\mid \mathcal{Z}_{i}, \mathbf{X}, \mathbf{U})$ and $\mathbf{p}=(p_{11},\dots, p_{Il_{I}})\in [0,1]^{L}$. A natural generalization of the Rosenbaum bounds (\ref{eqn: Rosenbaum bounds}) from matched to stratified observational studies \citep{rosenbaum2002observational, rosenbaum2018sensitivity} is: for some prespecified sensitivity parameter $\Upsilon$, we have 
\begin{equation*}
   \Upsilon^{-1} \leq p_{ik}/p_{ik^{\prime}}\leq \Upsilon \text{ for all $i\in \left\{1,\dots, I\right\}$ and $k, k^{\prime}\in \left\{1,\dots, l_{i} \right\}$}.  
\end{equation*}
Recall that $T=\sum_{i=1}^{I}\sum_{j=1}^{n_{i}}Z_{ij}q_{ij}$ denotes some sum statistic (e.g., permutational $t$-test, the Wilcoxon rank sum test, or $U$-statistics), where $q_{ij}$ is an arbitrary but prespecified score function based on the outcome data $\mathbf{Y}=(Y_{11},\dots, Y_{In_{I}})$. Then, recall that $S=\sum_{i=1}^{I}\sum_{j=1}^{n_{i}}Z_{ij}s_{ij}$ is some prespecified balance test statistic (see Section~\ref{sec: methods}), where each $s_{ij}$ is invariant under different $\mathbf{Z}$. Also, recall that $\mathcal{Z}_{i}=\{\mathbf{z}_{i}=(z_{i1},\dots, z_{in_{i}})\in \{0,1\}^{n_{i}}: \sum_{j=1}^{n_{i}}z_{ij}=m_{i}\}=\{\mathbf{b}_{i1}, \dots, \mathbf{b}_{il_{i}}\}$, $i=1,\dots, I$. For each $k\in \{1,\dots, l_{i}\}$, we let $\widetilde{q}_{ik}=\sum_{j=1}^{n_{i}}b_{ikj}q_{ij}$ represent the outcome test score contributed by stratum $i$ under $\mathbf{Z}_{i}=\mathbf{b}_{ik}$, and let $\widetilde{s}_{ik}=\sum_{j=1}^{n_{i}}b_{ikj}s_{ij}$ represent the balance test score contributed by stratum $i$ under $\mathbf{Z}_{i}=\mathbf{b}_{ik}$, respectively. Under the potential outcomes framework \citep{neyman1923application, rubin1974estimating}, the observed outcome for unit $j$ in stratum $i$ (denoted as $Y_{ij}$) satisfies $Y_{ij}=Z_{ij}Y_{ij}(1)+(1-Z_{ij})Y_{ij}(0)$, where $Y_{ij}(1)$ (or $Y_{ij}(0)$) denotes the corresponding potential outcome under treatment (or under control). Then, under $H_{0}: Y_{ij}(1)=Y_{ij}(0)$ for all $i$ and $j$, we have $E_{\mathbf{p}}(T)=\sum_{i=1}^{I}\sum_{k=1}^{l_{i}}p_{ik}\widetilde{q}_{ik}$, $\text{var}_{\mathbf{p}}(T) =\sum_{i=1}^{I}\big\{ \sum_{k=1}^{l_i}p_{ik}\widetilde{q}_{ik}^2-(\sum_{k=1}^{l_i}p_{ik}\widetilde{q}_{ik})^2
    \big\}$, $E_{\mathbf{p}}(S)=\sum_{i=1}^{I}\sum_{k=1}^{l_{i}}p_{ik}\widetilde{s}_{ik}$, and $\text{var}_{\mathbf{p}}(S) =\sum_{i=1}^{I}\big\{ \sum_{k=1}^{l_i}p_{ik}\widetilde{s}_{ik}^2-(\sum_{k=1}^{l_i}p_{ik}\widetilde{s}_{ik})^2
    \big\}$. Given any $c>0$, consider the following quadratically constrained quadratic program:
\begin{equation*}
     \begin{split}
        \underset{\mathbf{p}}{\text{minimize}} \quad & \{T-E_{\mathbf{p}}(T)\}^{2}-c\times \text{var}_{\mathbf{p}}(T) \quad \quad (\widetilde{L}_{c}) \\
         \text{subject to}\quad &\sum_{k=1}^{l_{i}} p_{ik}=1, \quad\forall i\in \{1,\dots, I\},\\
        &0 \leq p_{ik}\leq 1, \quad \forall i\in \{1,\dots, I\}, \forall k\in \{1,\dots, l_{i}\},\\
        & p_{ik}-\Upsilon\cdot p_{ik^{\prime}}\leq 0, \quad \forall i\in \{1,\dots, I\},\forall k, k^{\prime}\in  \{1,\dots, l_{i}\},\\
        & \{S-E_{\mathbf{p}}(S)\}^{2}-\chi_{1-\alpha^{\prime}, 1}^{2}\text{var}_{\mathbf{p}}(S) \leq 0.
     \end{split}
 \end{equation*}
By the same argument used in the proof of Theorem~\ref{thm: convexity}, the optimization problem $(\widetilde{L}_c)$ is convex. Moreover, following arguments analogous to those in Theorems~\ref{thm: validity} and \ref{thm: convergence} (with $q_{ij}$ and $s_{ij}$ replaced by $\widetilde{q}_{ik}$ and $\widetilde s_{ik}$), we can show that substituting $(\widetilde{L}_c)$ for $(L_c)$ in Algorithm~\ref{alg: part with covariate adjustment} yields an asymptotically valid worst-case $p$-value $\widetilde{p}^{*}$ for testing Fisher's sharp null $H_{0}$ in stratified observational studies. Importantly, $\widetilde{p}^{*}$ is more informative than the conventional sensitivity analysis $p$-value that ignores post-stratification overt bias (i.e., omits the constraint $\{S-E_{\mathbf{p}}(S)\}^2-\chi^2_{1-\alpha',1}\operatorname{var}_{\mathbf{p}}(S)\leq 0$).

Although this establishes that, in principle, the proposed iterative convex programming framework can be extended from matched to stratified observational studies, the extension raises additional computational challenges. In particular, the dimension of the decision variable in $(\widetilde L_c)$ is $\sum_{i=1}^I l_i=\sum_{i=1}^I{n_i\choose m_i}$, which can be prohibitively
large when at least one stratum contains many treated and control units. To address this issue, \citet{rosenbaum2018sensitivity} proposed combining a one-step Taylor approximation with the asymptotic separability algorithm of \citet{gastwirth2000asymptotic} to obtain statistically valid and computationally feasible bounds on worst-case $p$-values in sensitivity analyses for stratified observational studies. However, this approximation-based sensitivity analysis approach does not incorporate post-stratification overt bias information and is not readily adaptable to accommodate such information. Therefore, developing scalable methods that address the high dimensionality of $(\widetilde L_c)$ while incorporating overt bias constraints represents an important direction for future research.

\subsection{Generalizations to Neyman's Weak Null Hypotheses}

In the previous sections, we developed a more powerful sensitivity analysis framework for testing Fisher's sharp null hypotheses in matched observational studies. As mentioned in Section~\ref{subsec: rosenbaum bounds}, beyond Fisher's sharp null, another fundamental class of causal null hypotheses is Neyman's weak null hypotheses, which take the general form $H_{N,\tau_{0}}:\tau=\tau_{0}$, where $\tau = N^{-1}\sum_{i=1}^{I}\sum_{j=1}^{n_{i}}\{Y_{ij}(1)-Y_{ij}(0)\}$ denotes the sample average treatment effect and $\tau_{0}$ is a hypothesized value (e.g., $\tau_{0}=0$). In matched observational studies, two main approaches have been developed for sensitivity analysis under the Rosenbaum-type sensitivity analysis framework when testing Neyman's weak null hypotheses. The first is an \emph{analytical approach} proposed by \citet{fogarty2020studentized, fogarty2023testing}, which yields explicit asymptotic solutions for calculating the worst-case (maximal) $p$-value. The second is an \emph{optimization-based approach} developed in Chapter~5 of \citet{fogarty2016modern}, which formulates the computation of the worst-case $p$-value as an optimization problem amenable to numerical solution. In the special case of pair-matched observational studies, in which each matched pair contains one treated and one control unit, \citet{fogarty2016modern} further proposed a specialized optimization formulation that can be solved in linear time. Numerical investigations in \citet{fogarty2016modern, fogarty2020studentized} suggest that, in pair-matched observational studies, the optimization-based approach is typically more conservative than the analytical approach. Nevertheless, the optimization-based framework offers greater flexibility, particularly in its potential to incorporate post-matching overt bias information to improve power. Motivated by this observation, we discuss how the proposed confounder-imbalance-constrained sensitivity analysis framework may be extended to strengthen the optimization-based approach for testing $H_{N,\tau_{0}}:\tau=\tau_{0}$, along with the associated computational challenges.

Consider a pair-matched observational study with $I$ independent pairs. Following \citet{fogarty2016modern}, for notational simplicity, we assume that the observed treatment assignments satisfy $Z_{i1}=1$ and $Z_{i2}=0$ for all matched pairs $i$; otherwise, the indices $i1$ and $i2$ can be relabeled. Consider the Neyman (difference-in-means) estimator for $\tau$: $T_{N}=I^{-1}\sum_{i=1}^{I}(Z_{i1}-Z_{i2})(Y_{i1}-Y_{i2})$. Let $\varphi_{i1}=Y_{i1}(1)-Y_{i2}(0)=Y_{i1}-Y_{i2}$ and $\varphi_{i2}=Y_{i2}(1)-Y_{i1}(0)$, and define $\boldsymbol{\pi}=(\pi_{1},\dots, \pi_{I})$ with each $\pi_i = p_{i1} = \text{pr}(Z_{i1}=1 \mid \mathcal{Z}, \mathbf{X}, \mathbf{U})$. Then, under this setup, the observed value of $T_{N}$ is $I^{-1}\sum_{i=1}^{I}\varphi_{i1}$, and we have $E_{\boldsymbol{\pi}}(T_{N})=I^{-1}\sum_{i=1}^{I}\left\{\pi_{i}\varphi_{i1}+(1-\pi_{i})\varphi_{i2}\right\}$ and $\text{var}_{\boldsymbol{\pi}}(T_{N})=I^{-2}\sum_{i=1}^{I} \pi_{i}(1-\pi_{i})(\varphi_{i1}-\varphi_{i2})^{2}$. Then, the $z$-score for $T_{N}$ is $\overline{T}_{N,\boldsymbol{\pi}}=\frac{T_{N}-E_{\boldsymbol{\pi}}(T_{N})}{\sqrt{\text{var}_{\boldsymbol{\pi}}(T_{N})}}= \frac{\sum_{i=1}^{I}\varphi_{i1}-\sum_{i=1}^{I}\left\{\pi_{i}\varphi_{i1}+(1-\pi_{i})\varphi_{i2}\right\} }{\sqrt{\sum_{i=1}^{I} \pi_{i}(1-\pi_{i})(\varphi_{i1}-\varphi_{i2})^{2}}}$. By the finite-population central limit theorem \citep{li2017general}, to find the worst-case (maximal) $p$-value (one-sided, greater than) is equivalent to finding the worst-case (minimal) $z$-score, as proposed in \citet{fogarty2016modern}. Specifically, when $\Gamma>1$, the optimization problem (P1) in Chapter 5 of \citet{fogarty2016modern} proposes to minimize the $z$-score $\overline{T}_{N,\boldsymbol{\pi}}=\frac{\sum_{i=1}^{I}\varphi_{i1}-\sum_{i=1}^{I}\left\{\pi_{i}\varphi_{i1}+(1-\pi_{i})\varphi_{i2}\right\} }{\sqrt{\sum_{i=1}^{I} \pi_{i}(1-\pi_{i})(\varphi_{i1}-\varphi_{i2})^{2}}}$ subject to the following three constraints: (i) $\sum_{i=1}^{I} \{\varphi_{i1}+\varphi_{i2}\}=2I \tau_{0}$, which holds under $H_{N, \tau_{0}}$; (ii) $\sum_{i=1}^{I}\pi_{i}(1-\pi_{i})(\varphi_{i1}-\varphi_{i2})^{2}\leq  \frac{2\Gamma}{1+\Gamma}\sum_{i=1}^{I}(\varphi_{i1}-\tau_{0})^{2}$, which, under the true $\mathbf{p}$ and each fixed $\Gamma>1$, as well as corresponding regularity conditions, holds with probability tending to one as $I\to\infty$,  by applying Chebyshev's inequality to the variance estimator in Proposition 2 in Chapter 5 of \citet{fogarty2016modern}; and (iii) $\frac{1}{1+\Gamma}\leq  \pi_{i} \leq \frac{\Gamma}{1+\Gamma}$ for all $i=1,\dots, I$, which is an equivalent form of the Rosenbaum bounds constraint (\ref{eqn: Rosenbaum bounds}) in pair-matched design. 

Following the proposal in Section~\ref{sec: methods} of this manuscript, we consider a balance test statistic $S=\sum_{i=1}^{I}\sum_{j=1}^{2}Z_{ij}s_{ij}$. Under the above setup, the observed value of $S$ is $\sum_{i=1}^{I}s_{i1}$, and we have $E_{\boldsymbol{\pi}}(S)=\sum_{i=1}^{I}\left\{\pi_{i}s_{i1}+(1-\pi_{i})s_{i2}\right\}$ and $\text{var}_{\boldsymbol{\pi}}(S)=\sum_{i=1}^{I} \pi_{i}(1-\pi_{i})(s_{i1}-s_{i2})^{2}$. By augmenting the optimization problem (P1) in Chapter~5 of \citet{fogarty2016modern} with an additional constraint that incorporates post-matching confounder imbalance information proposed in Section~\ref{sec: methods} (i.e., the constraint $\{S-E_{\boldsymbol{\pi}}(S)\}^{2}-\chi_{1-\alpha^{\prime}, 1}^{2}\text{var}_{\boldsymbol{\pi}}(S) \leq 0$), we consider the following optimization problem:
\begin{equation*}
     \begin{split}
        \underset{\pi_{i}, \varphi_{i2}}{\text{minimize}} \quad  & \frac{\sum_{i=1}^{I}\varphi_{i1}-\sum_{i=1}^{I}\left\{\pi_{i}\varphi_{i1}+(1-\pi_{i})\varphi_{i2}\right\} }{\sqrt{\sum_{i=1}^{I} \pi_{i}(1-\pi_{i})(\varphi_{i1}-\varphi_{i2})^{2}}} \quad \quad (O_{\tau_{0}})\\
         \text{subject to}\quad & \sum_{i=1}^{I} \varphi_{i1}+\varphi_{i2}=2I \tau_{0},  \\
        &\sum_{i=1}^{I}\pi_{i}(1-\pi_{i})(\varphi_{i1}-\varphi_{i2})^{2}\leq  \frac{2\Gamma}{1+\Gamma}\sum_{i=1}^{I}(\varphi_{i1}-\tau_{0})^{2},  \\ 
         & \frac{1}{1+\Gamma}\leq  \pi_{i} \leq \frac{\Gamma}{1+\Gamma}, \quad \forall i\in \{1,\dots, I\}, \\
         & \left[\sum_{i=1}^{I}s_{i1}-\sum_{i=1}^{I}\left\{\pi_{i}s_{i1}+(1-\pi_{i})s_{i2}\right\}\right]^{2}-\chi^{2}_{1-\alpha^{\prime},1} \sum_{i=1}^{I} \pi_{i}(1-\pi_{i})(s_{i1}-s_{i2})^{2}\leq 0.
     \end{split}
\end{equation*}
Let $\widetilde{d}^{*}$ denote the optimal value of $(O_{\tau_{0}})$. Combining the arguments in \citet{fogarty2016modern} with a multiplicity adjustment for the balance test $S$ and the outcome test $T_{N}$ (i.e., similar to Step 3 in Section~\ref{sec: methods}), and using reasoning analogous to the proof of Theorem~\ref{thm: validity}, the quantity $\widetilde{p}^{*}_{N}=\{1-\Phi(\widetilde{d}^{*})\}+\alpha^{\prime}$ constitutes a valid one-sided $p$-value for sensitivity analysis under Neyman's weak null hypothesis $H_{N,\tau_{0}}$. However, unlike the optimization problems $(L_{c})$ associated with Fisher's sharp null hypotheses, the optimization problem $(O_{\tau_{0}})$ is non-convex. Developing computationally efficient algorithms (exact or approximate) for solving $(O_{\tau_{0}})$, therefore, represents a meaningful direction for future research.

\section*{Supplementary Materials}

The online supplementary materials contain detailed proofs, additional simulation studies, and further discussions and remarks.

\section*{Declaration of Generative AI Use}

The authors used generative AI tools to assist with proofreading for grammatical, typographical, and notational inconsistencies; reviewing the logical consistency and improving the rigor and clarity of mathematical proofs; and drafting some basic functions in \textsf{R} and preparing accompanying \textsf{R} Markdown files. The authors confirm the originality and integrity of the manuscript and take full responsibility for its content, including the proposed methods, mathematical proofs, computational code, and theoretical and numerical results.

\section*{Acknowledgments}

The authors thank the Joint Editor, the Associate Editor, and two anonymous reviewers for their helpful comments and insightful suggestions, which greatly improved the manuscript.

\putbib[references]  
\end{bibunit}

\clearpage

\spacingset{1.73}

\begin{bibunit}[apalike]

\begin{center}
    \Large \bf Supplementary Materials for ``Reconciling Overt Bias and Hidden Bias in Sensitivity Analysis for Matched Observational Studies'' 
\end{center}

\begin{abstract}
The supplementary materials provide supporting details, additional results, and further discussions. Appendix A presents the technical proofs for the main theoretical results. Appendix B provides additional details and remarks that complement the main text. Appendix C reports the computational running times of the proposed sensitivity analysis approach. Appendix D discusses an extension to multivariate balance test scores. Appendix E presents preliminary investigations and practical guidance for constructing balance test scores, including initial insights on how to incorporate additional knowledge about potential outcomes. Appendix F investigates the performance of the proposed approach when external data are used to estimate propensity scores. Appendix G describes the additional simulation results under different standardized-difference thresholds. Appendix H concludes with a brief summary of our work.
\end{abstract}

\section*{Appendix A: Proofs}

\subsection*{A.1: Proof of Theorem~\ref{thm: validity}}

\begin{proof}
    
Fix $\Gamma\geq 1$ and $0<\alpha'<\alpha<1$. Recall that $\mathbf{p}=(p_{11},\dots, p_{In_{I}})\in \Lambda_{\Gamma}$ represent the true post-matching treatment assignment probabilities. To define the testing procedure for every possible treatment assignment, we set the $p$-value $p^{*}=1$ whenever $\Lambda_{\Gamma}\cap\Delta_{1-\alpha^{\prime}}=\emptyset$, corresponding to nonrejection at this fixed $\Gamma$; otherwise, $p^{*}$ is defined by the formula in the main text. In the rest of the proof, we only need to justify the validity of $p^{*}$ when $\Lambda_{\Gamma}\cap \Delta_{1-\alpha^{\prime}}\neq \emptyset$.

Note that when $\mathbf{p}\in \Lambda_{\Gamma}\cap \Delta_{1-\alpha^{\prime}}$, we have 
\begin{equation*}
d_{\mathbf{p}}:=\frac{\{T-E_{\mathbf{p} }(T)\}^{2}}{\text{var}_{\mathbf{p}}(T)}\geq \min_{\mathbf{p}^{\prime}\in \Lambda_{\Gamma}\cap \Delta_{1-\alpha^{\prime}} }\frac{\{T-E_{\mathbf{p}^{\prime} }(T)\}^{2}}{\text{var}_{\mathbf{p}^{\prime}}(T)}=d^{*}.
\end{equation*}
By Assumptions~\ref{assump: independence assumption} and \ref{assump: CLT for T}, by the Lindeberg-Feller central limit theorem, we have 
\begin{equation}\label{eqn: asymptotic normality of T}
  \frac{T-E_{\mathbf{p} }(T)}{\sqrt{\text{var}_{\mathbf{p}}(T)}}\xrightarrow{d} N(0,1) \quad \text{as $I\rightarrow\infty$}.
\end{equation}
Also, under $H_{0}$, we have 
\begin{align}\label{eqn: event relationship}
   \left \{p^{*}\leq \alpha, \mathbf{p}\in \Delta_{1-\alpha^{\prime}}\right\}&=\left\{  2\{1-\Phi(\sqrt{d^{*}})\}+\alpha^{\prime}\leq \alpha, \mathbf{p}\in \Delta_{1-\alpha^{\prime}}  \right\}\nonumber \\
   &\subseteq \left\{ 2\{1-\Phi(\sqrt{d_{\mathbf{p}} })\}+\alpha^{\prime}\leq \alpha, \mathbf{p}\in \Delta_{1-\alpha^{\prime}}  \right\}\nonumber \\
   &\subseteq  \left\{ 2\{1-\Phi(\sqrt{d_{\mathbf{p}} })\}+\alpha^{\prime}\leq \alpha \right\}.
\end{align}
By (\ref{eqn: asymptotic normality of T}) and (\ref{eqn: event relationship}), we have 
\begin{equation}\label{eqn: subset event prob}
  \limsup_{I\rightarrow \infty} \text{pr}_{\mathbf{p}}(p^{*}\leq \alpha, \mathbf{p}\in \Delta_{1-\alpha^{\prime}} \mid H_{0})\leq  \limsup_{I\rightarrow \infty} \text{pr}_{\mathbf{p}}\big( 2\{1-\Phi(\sqrt{d_{\mathbf{p}} })\}+\alpha^{\prime}\leq \alpha \mid H_{0}\big)=\alpha-\alpha^{\prime}.
\end{equation}
Under Assumptions~\ref{assump: independence assumption} and \ref{assump: CLT for S}, by Lemma~\ref{thm: validity of A}, we have 
\begin{align}\label{eqn: probability of not contain}
    \limsup_{I\rightarrow \infty} \text{pr}_{\mathbf{p}}(\mathbf{p}\notin \Delta_{1-\alpha^{\prime}} \mid H_{0})&=1-\liminf_{I\rightarrow \infty} \text{pr}_{\mathbf{p}}(\mathbf{p}\in \Delta_{1-\alpha^{\prime}} \mid H_{0})\nonumber \\
    &=1-\lim_{I\rightarrow \infty} \text{pr}_{\mathbf{p}}(\mathbf{p}\in \Delta_{1-\alpha^{\prime}} \mid H_{0})\nonumber \\
    &=\alpha^{\prime}. 
\end{align}
By (\ref{eqn: subset event prob}) and (\ref{eqn: probability of not contain}), we have
\begin{align*}
 \limsup_{I\rightarrow \infty} \text{pr}_{\mathbf{p}}(p^{*}\leq \alpha \mid H_{0})
 &\leq \limsup_{I\rightarrow \infty}\text{pr}_{\mathbf{p}}(p^{*}\leq \alpha, \mathbf{p}\in \Delta_{1-\alpha^{\prime}} \mid H_{0})+ \limsup_{I\rightarrow \infty} \text{pr}_{\mathbf{p}}(p^{*}\leq \alpha, \mathbf{p}\notin \Delta_{1-\alpha^{\prime}} \mid H_{0})\\
    &\leq  \limsup_{I\rightarrow \infty} \text{pr}_{\mathbf{p}}(p^{*}\leq \alpha, \mathbf{p}\in \Delta_{1-\alpha^{\prime}} \mid H_{0})+ \limsup_{I\rightarrow \infty} \text{pr}_{\mathbf{p}}(\mathbf{p}\notin \Delta_{1-\alpha^{\prime}} \mid H_{0})\\
    &\leq \alpha-\alpha^{\prime}+\alpha^{\prime}=\alpha.
\end{align*}

\end{proof}

\subsection*{A.2: Proof of Theorem~\ref{thm: convexity}}

As specified in Algorithm~\ref{alg: part with covariate adjustment}, we proceed to outcome analysis (including sensitivity analysis) at a prespecified sensitivity parameter $\Gamma$ only if $\Gamma$ is not falsified by the balance test $S$, i.e., when $\Lambda_{\Gamma}\cap\Delta_{1-\alpha^{\prime}}\neq\emptyset$. Accordingly, in Theorems~\ref{thm: convexity} and \ref{thm: convergence} and their corresponding proofs in Appendices A.2 and A.3, we consider the case where $\Lambda_{\Gamma}\cap\Delta_{1-\alpha^{\prime}} \neq\emptyset$. Otherwise, Algorithm~\ref{alg: part with covariate adjustment} increases $\Gamma$ until the feasible set $\Lambda_{\Gamma}\cap\Delta_{1-\alpha^{\prime}} \neq \emptyset$.

\begin{proof}
Let $\mathcal{I}_{1}=\{i\in \{1,\dots, I\}: m_{i}=1\}$ and $\mathcal{I}_{2}=\{i\in \{1,\dots, I\}: n_{i}-m_{i}=1 \text{ and } m_{i}\geq 2\}$. We have
    \begin{align*}
        g(\mathbf{p})&=\{T-E_{\mathbf{p}}(T)\}^{2}-c\times \text{var}_{\mathbf{p}}(T)\\
        &=\Big\{T-\sum_{i\in \mathcal{I}_{1}}\sum_{j=1}^{n_{i}}p_{ij}q_{ij}-\sum_{i\in \mathcal{I}_{2}}\sum_{j=1}^{n_{i}}(1-p_{ij})q_{ij} \Big\}^{2}-c\times \sum_{i=1 }^{I}\Big\{\sum_{j=1}^{n_i}p_{ij}q_{ij}^2-\Big(\sum_{j=1}^{n_i}p_{ij}q_{ij}\Big)^2\Big\}\\
        &=\sum_{i=1}^{I}\sum_{j=1}^{n_{i}}\sum_{i^{\prime}=1}^{I}\sum_{j^{\prime}=1}^{n_{i^{\prime}}}a_{iji^{\prime}j^{\prime}}p_{ij}p_{i^{\prime}j^{\prime}}+\sum_{i=1}^{I}\sum_{j = 1}^{n_{i}}b_{ij}p_{ij}+\Big(T-\sum_{i\in \mathcal{I}_{2}}\sum_{j=1}^{n_{i}}q_{ij}\Big)^{2},
    \end{align*}
    where 
    \begin{equation*}
a_{iji^{\prime}j^{\prime}}=\left\{
\begin{array}{ll}
       &q_{ij}q_{i^{\prime}j^{\prime}}(1+c\times \delta_{ii^{\prime}}) \quad \text{for\ $i\in \mathcal{I}_{1}, i^{\prime}\in \mathcal{I}_{1}$ or $i\in \mathcal{I}_{2}, i^{\prime}\in \mathcal{I}_{2}$}, \\
        & -q_{ij}q_{i^{\prime}j^{\prime}}  \quad \text{for\ $i\in \mathcal{I}_{1}, i^{\prime}\in \mathcal{I}_{2}$ or $i\in \mathcal{I}_{2}, i^{\prime }\in \mathcal{I}_{1}$}, 
\end{array}
\right.
\end{equation*}
and
\begin{equation*}
b_{ij}=\left\{
\begin{array}{ll}
       &-2\big(T-\sum_{i^{\prime}\in \mathcal{I}_{2}}\sum_{j^{\prime}=1}^{n_{i^{\prime}}}q_{i^{\prime}j^{\prime}}\big)q_{ij}-c\times q_{ij}^{2} \quad \text{for\ $i\in \mathcal{I}_{1}$}, \\
        &2\big(T-\sum_{i^{\prime}\in \mathcal{I}_{2}}\sum_{j^{\prime}=1}^{n_{i^{\prime}}}q_{i^{\prime}j^{\prime}}\big)q_{ij}-c\times q_{ij}^{2}  \quad \text{for\ $i\in \mathcal{I}_{2}$}, 
\end{array}
\right.
\end{equation*}
in which $\delta_{ii^{\prime}}=\mathbbm{1}\{i=i^{\prime}\}$. Then, we can write $g(\mathbf{p})=\mathbf{p}\mathbf{A}\mathbf{p}^{T}+\mathbf{b}\mathbf{p}^{T}+(T-\sum_{i\in \mathcal{I}_{2}}\sum_{j=1}^{n_{i}}q_{ij})^{2}$, where $\mathbf{A}$ is an $N\times N$ matrix (here $N=\sum_{i=1}^{I}n_{i}$ denotes the total number of units), of which the $N$ rows and columns of $\mathbf{A}$ are indexed by $ij$ as $i$ runs over the $I$ matched sets and $j$ runs over $n_{i}$ units in each matched set $i$, and $\mathbf{b}=(b_{11},\dots, b_{In_{I}})$. To show that $g(\mathbf{p})$ is convex, it suffices to show that $\mathbf{A}$ is positive semidefinite. Let $\mathbf{v}=\{v_{ij}: i=1,\dots, I, j=1,\dots, n_{i}\}$ be any $N$-dimensional vector. Then, for any $c\geq 0$, we have 
\begin{align*}
    \mathbf{v}\mathbf{A}\mathbf{v}^{T}&=\sum_{i=1}^{I}\sum_{j=1}^{n_{i}}\sum_{i^{\prime}=1}^{I}\sum_{j^{\prime}=1}^{n_{i^{\prime}}}a_{iji^{\prime}j^{\prime}}v_{ij}v_{i^{\prime}j^{\prime}}\\
    &= \Big(\sum_{i\in \mathcal{I}_{1} }\sum_{j=1}^{n_{i }}v_{ij}q_{ij}-\sum_{i\in \mathcal{I}_{2}}\sum_{j=1}^{n_{i }}v_{ij}q_{ij}\Big)^{2}+ c\times  \sum_{i=1}^{I}\Big(\sum_{j=1}^{n_{i }}v_{ij}q_{ij}\Big)^{2}\geq 0.
\end{align*}
Therefore, the matrix $\mathbf{A}$ is positive semidefinite, and $g(\mathbf{p})$ is a convex function. 

Similarly, we have
    \begin{align*}
        h(\mathbf{p})&=\{S-E_{\mathbf{p}}(S)\}^{2}-\chi_{1-\alpha^{\prime}, 1}^{2}\times \text{var}_{\mathbf{p}}(S)\\
        &=\Big\{S-\sum_{i\in \mathcal{I}_{1}}\sum_{j=1}^{n_{i}}p_{ij}s_{ij}-\sum_{i\in \mathcal{I}_{2}}\sum_{j=1}^{n_{i}}(1-p_{ij})s_{ij} \Big\}^{2}-\chi_{1-\alpha^{\prime}, 1}^{2}\times \sum_{i=1 }^{I}\Big\{\sum_{j=1}^{n_i}p_{ij}s_{ij}^2-\Big(\sum_{j=1}^{n_i}p_{ij}s_{ij}\Big)^2\Big\}\\
        &=\sum_{i=1}^{I}\sum_{j=1}^{n_{i}}\sum_{i^{\prime}=1}^{I}\sum_{j^{\prime}=1}^{n_{i^{\prime}}}\widetilde{a}_{iji^{\prime}j^{\prime}}p_{ij}p_{i^{\prime}j^{\prime}}+\sum_{i=1}^{I}\sum_{j = 1}^{n_{i}}\widetilde{b}_{ij}p_{ij}+\Big(S-\sum_{i\in \mathcal{I}_{2}}\sum_{j=1}^{n_{i}}s_{ij}\Big)^{2},
    \end{align*}
     where 
    \begin{equation*}
\widetilde{a}_{iji^{\prime}j^{\prime}}=\left\{
\begin{array}{ll}
       &s_{ij}s_{i^{\prime}j^{\prime}}(1+\chi_{1-\alpha^{\prime}, 1}^{2}\times \delta_{ii^{\prime}}) \quad \text{for\ $i\in \mathcal{I}_{1}, i^{\prime}\in \mathcal{I}_{1}$ or $i\in \mathcal{I}_{2}, i^{\prime}\in \mathcal{I}_{2}$}, \\
        & -s_{ij}s_{i^{\prime}j^{\prime}}  \quad \text{for\ $i\in \mathcal{I}_{1}, i^{\prime}\in \mathcal{I}_{2}$ or $i\in \mathcal{I}_{2}, i^{\prime }\in \mathcal{I}_{1}$}, 
\end{array}
\right.
\end{equation*}
and
\begin{equation*}
\widetilde{b}_{ij}=\left\{
\begin{array}{ll}
       &-2\big(S-\sum_{i^{\prime}\in \mathcal{I}_{2}}\sum_{j^{\prime}=1}^{n_{i^{\prime}}}s_{i^{\prime}j^{\prime}}\big)s_{ij}-\chi_{1-\alpha^{\prime}, 1}^{2}\times s_{ij}^{2} \quad \text{for\ $i\in \mathcal{I}_{1}$}, \\
        &2\big(S-\sum_{i^{\prime}\in \mathcal{I}_{2}}\sum_{j^{\prime}=1}^{n_{i^{\prime}}}s_{i^{\prime}j^{\prime}}\big)s_{ij}-\chi_{1-\alpha^{\prime}, 1}^{2}\times s_{ij}^{2}  \quad \text{for\ $i\in \mathcal{I}_{2}$}, 
\end{array}
\right.
\end{equation*}
in which $\delta_{ii^{\prime}}=\mathbbm{1}\{i=i^{\prime}\}$. Similarly, we can show that  $h(\mathbf{p})$ is also a convex function. Since the constraints $p_{ij}\geq 0$ and $\sum_{j=1}^{n_{i}}p_{ij}=1$ are linear constraints, and both $g(\mathbf{p})$ and $h(\mathbf{p})$ are convex functions, the quadratically constrained quadratic program (\ref{eqn: convex program}) considered in Theorem~\ref{thm: convexity} is convex for any fixed $c\geq 0$.

\end{proof}

\subsection*{A.3: Proof of Theorem~\ref{thm: convergence}}

\begin{lemma}\label{lem: var bounds}
     Under Assumptions~\ref{assump: independence assumption}--\ref{assump: CLT for T}, we have $v_{\min}=\min_{\mathbf{p}\in \Lambda_{\Gamma}\cap \Delta_{1-\alpha^{\prime}}}\text{var}_{\mathbf{p}}(T)>0$ and $v_{\max}=\max_{\mathbf{p}\in \Lambda_{\Gamma}\cap \Delta_{1-\alpha^{\prime}}}\text{var}_{\mathbf{p}}(T)<\infty$.
\end{lemma}

\begin{proof}
 Since $\max_{i}\max_{j, j^{\prime}}|q_{ij}-q_{ij^{\prime}}|>0$ and $\max_{i}\max_{j, j^{\prime}}|s_{ij}-s_{ij^{\prime}}|>0$, we have $\text{var}_{\mathbf{p}}(T)>0$ and $\text{var}_{\mathbf{p}}(S)>0$ for any $\mathbf{p}\in \Lambda_{\Gamma}$. Also, since $\{S-E_{\mathbf{p}}(S)\}^{2}/\text{var}_{\mathbf{p}}(S)$ is a continuous function of $\mathbf{p}$ over $\mathbf{p}\in \Lambda_{\Gamma}$, by the definition of $\Delta_{1-\alpha^{\prime}}$, we know that $ \Lambda_{\Gamma}\cap \Delta_{1-\alpha^{\prime}}$ is a compact set. This also implies that $\text{var}_{\mathbf{p}}(T)<\infty$ for any $\mathbf{p}\in \Lambda_{\Gamma}\cap \Delta_{1-\alpha^{\prime}}$. Therefore, since $ \Lambda_{\Gamma}\cap \Delta_{1-\alpha^{\prime}}$ is compact and $\text{var}_{\mathbf{p}}(T)$ is a continuous and positive function of $\mathbf{p}$ over $\mathbf{p}\in \Lambda_{\Gamma}\cap \Delta_{1-\alpha^{\prime}}$, we have $v_{\min}=\min_{\mathbf{p}\in \Lambda_{\Gamma}\cap \Delta_{1-\alpha^{\prime}}}\text{var}_{\mathbf{p}}(T)>0$ and $v_{\max}=\max_{\mathbf{p}\in \Lambda_{\Gamma}\cap \Delta_{1-\alpha^{\prime}}}\text{var}_{\mathbf{p}}(T)<\infty$. 
\end{proof}

The proof of Theorem~\ref{thm: convergence} follows convergence arguments for nonlinear fractional programs similar to those in \citet{dinkelbach1967nonlinear} and \citet{fogarty2016discrete}.

    \begin{proof}
Note that the sequence $\{c_{(0)}, c_{(1)}, c_{(2)}, \dots, \}$ is bounded below by zero. Also, for $m=1,2,\dots$, we have
\begin{equation*}
    \{T-E_{\mathbf{p}_{(m+1)}}(T)\}^{2}-c_{(m)}\text{var}_{\mathbf{p}_{(m+1)}}(T)\leq \{T-E_{\mathbf{p}_{(m)}}(T)\}^{2}-c_{(m)}\text{var}_{\mathbf{p}_{(m)}}(T)=0.
\end{equation*}
Therefore, for $m=1,2,\dots$, we have 
\begin{equation*}
     c_{(m+1)}=\{T-E_{\mathbf{p}_{(m+1)}}(T)\}^{2}/\text{var}_{\mathbf{p}_{(m+1)}}(T)\leq c_{(m)}.
\end{equation*}
By the monotone convergence theorem, the sequence $c_{(m)}$ converges to a limit $c^{*}$ as $m \rightarrow \infty$. We now show that the stationary point $c^{*}$ equals the optimal value $d^{*}$ of the nonlinear fractional program (\ref{eqn: fractional program}) stated in the main text. 

We first show that $c^{*}\geq d^{*}$. Note that for every $m\ge1$, we have $\mathbf p_{(m)}\in  \Lambda_{\Gamma}\cap \Delta_{1-\alpha^{\prime}}$. Therefore, we have $c_{(m)}=\{T-E_{\mathbf{p}_{(m)}}(T)\}^{2}/\text{var}_{\mathbf{p}_{(m)}}(T)\geq d^{*}$ for every $m\geq 1$. This implies that $c^{*}=\lim_{m\rightarrow \infty}c_{(m)} \geq d^{*}$.

Then, we show that $c^{*}\leq d^{*}$. Let $\mathbf{p}^{*}\in \Lambda_{\Gamma}\cap \Delta_{1-\alpha^{\prime}}$ denote the optimal solution to the nonlinear fractional program (\ref{eqn: fractional program}). Note that 
\begin{align}\label{eqn: var product inequality}
     \text{var}_{\mathbf{p}_{(m+1)}}(T)\cdot \left(c_{(m+1)}-c_{(m)}\right)&= \{T-E_{\mathbf{p}_{(m+1)}}(T)\}^{2}-c_{(m)}\cdot \text{var}_{\mathbf{p}_{(m+1)}}(T)\nonumber \\
     &\leq \{T-E_{\mathbf{p}^{*}}(T)\}^{2}-c_{(m)}\cdot \text{var}_{\mathbf{p}^{*}}(T) \nonumber \\
     &=\text{var}_{\mathbf{p}^{*}}(T)\cdot \left(d^{*}-c_{(m)}\right).
\end{align}
Since we have shown that the sequence $c_{(m)}$ converges to a limit $c^{*}$ as $m \rightarrow \infty$, we have $c_{(m+1)}-c_{(m)}\rightarrow 0$ as $m\rightarrow \infty$. Also, we have $\text{var}_{\mathbf{p}_{(m+1)}}(T)\leq v_{\text{max}}<+\infty$ (by Lemma~\ref{lem: var bounds}). Combining the above two facts, by (\ref{eqn: var product inequality}), we have
\begin{align*}
    \text{var}_{\mathbf{p}^{*}}(T)\cdot \left(d^{*}-c^{*}\right) =\lim_{m\rightarrow \infty} \text{var}_{\mathbf{p}^{*}}(T)\cdot \left(d^{*}-c_{(m)}\right) \geq \lim_{m\rightarrow \infty}\text{var}_{\mathbf{p}_{(m+1)}}(T)\cdot \left(c_{(m+1)}-c_{(m)}\right)=0.
\end{align*}
Because $\text{var}_{\mathbf{p}^{*}}(T)\geq v_{\text{min}}>0$ (by Lemma~\ref{lem: var bounds}) and $\text{var}_{\mathbf{p}^{*}}(T)\cdot \left(d^{*}-c^{*}\right)\geq 0$, we have $c^{*}\leq d^{*}$.

Putting the above arguments together, we have shown that $\lim_{m\rightarrow \infty}c_{(m)}=c^{*}=d^{*}$, which implies that $\lim_{m\rightarrow \infty}p_{(m)}= p^{*}=2\{1-\Phi(\sqrt{d^{*}})\}+\alpha^{\prime}$.
\end{proof}

\begin{remark}
In the proof of Theorem~\ref{thm: convergence}, we showed that $c^{*}=d^{*}$ by establishing both $c^{*}\geq d^{*}$ and $c^{*}\leq d^{*}$. Recall that $d^{*}$ is the worst-case (minimal) squared deviate over $\mathbf{p}\in \Lambda_{\Gamma}\cap \Delta_{1-\alpha^{\prime}}$, that is, the optimal value of the quadratic fractional program~(\ref{eqn: fractional program}). Because $d^{*}$ does not depend on the starting value $c_{(0)}$, the stationary point $c^{*}$ is also independent of $c_{(0)}$.
\end{remark}

\section*{Appendix B: Additional Details and Remarks}

\begin{remark}\label{rem: formulas of pij}
   Recall that $e_{ij}=\text{pr}(Z_{ij}=1\mid \mathbf{x}_{ij}, \mathbf{u}_{ij})$ denotes the true (pre-matching) propensity score of unit $j$ in matched set $i$ (possibly in the presence of unmeasured confounders). Following the arguments in \citet{rosenbaum1988permutation}, \citet{pimentel2024covariate}, and \citet{zhu2025randomization}, if $m_{i}=\sum_{j=1}^{n_{i}}Z_{ij}=1$, we have 
\begin{align}\label{eqn: pij}
    p_{ij}&=\text{pr}(Z_{i1}=0, \dots, Z_{i(j-1)}=0, Z_{ij}=1, Z_{i(j+1)}=0, \dots, Z_{in_{i}}=0 \mid m_{i}=1, \mathbf{X}, \mathbf{U})\nonumber \\
    &=\frac{\text{pr}(Z_{i1}=0, \dots, Z_{i(j-1)}=0, Z_{ij}=1, Z_{i(j+1)}=0, \dots, Z_{in_{i}}=0 \mid \mathbf{X}, \mathbf{U})}{\sum_{j^{\prime}=1}^{n_{i}}\text{pr}(Z_{i1}=0, \dots, Z_{i(j^{\prime}-1)}=0, Z_{ij^{\prime}}=1, Z_{i(j^{\prime}+1)}=0, \dots, Z_{in_{i}}=0 \mid \mathbf{X}, \mathbf{U}) }\nonumber \\
    &=\frac{e_{ij}\prod_{1\leq \tilde{j}\leq n_{i}, \tilde{j}\neq j}(1-e_{i\tilde{j}})}{\sum_{j^{\prime}=1}^{n_{i}}\{e_{ij^{\prime}}\prod_{1\leq \tilde{j}\leq n_{i}, \tilde{j}\neq j^{\prime} }(1-e_{i\tilde{j}})\} }.
\end{align}
If $n_{i}-m_{i}=\sum_{j=1}^{n_{i}}(1-Z_{ij})=1$ and $m_{i}\geq 2$, we have 
\begin{align}\label{eqn: qij}
    p_{ij}&=\text{pr}(Z_{i1}=1, \dots, Z_{i(j-1)}=1, Z_{ij}=0, Z_{i(j+1)}=1, \dots, Z_{in_{i}}=1 \mid n_{i}-m_{i}=1, \mathbf{X}, \mathbf{U})\nonumber \\
    &=\frac{\text{pr}(Z_{i1}=1, \dots, Z_{i(j-1)}=1, Z_{ij}=0, Z_{i(j+1)}=1, \dots, Z_{in_{i}}=1 \mid \mathbf{X}, \mathbf{U})}{\sum_{j^{\prime}=1}^{n_{i}}\text{pr}(Z_{i1}=1, \dots, Z_{i(j^{\prime}-1)}=1, Z_{ij^{\prime}}=0, Z_{i(j^{\prime}+1)}=1, \dots, Z_{in_{i}}=1 \mid \mathbf{X}, \mathbf{U}) }\nonumber \\
    &=\frac{(1-e_{ij})\prod_{1\leq \tilde{j}\leq n_{i}, \tilde{j}\neq j}e_{i\tilde{j}}}{\sum_{j^{\prime}=1}^{n_{i}}\{(1-e_{ij^{\prime}})\prod_{1\leq \tilde{j}\leq n_{i},  \tilde{j}\neq j^{\prime} }e_{i\tilde{j}}\} }.
\end{align}
Therefore, to obtain the estimated value $\widehat{p}_{ij}$ of each $p_{ij}$, we can first obtain the estimated propensity scores $\widehat{\mathbf{e}}=(\widehat{e}_{11}, \dots, \widehat{e}_{In_{I}})$ and then plug them in formulas (\ref{eqn: pij}) and (\ref{eqn: qij}). 
\end{remark}

\begin{remark}
When implementing the program $(L_{c})$, we can replace the box constraints $0\leq p_{ij}\leq 1$ with the more informative box constraints $\frac{1}{1+(n_{i}-1)\Gamma} \leq p_{ij}\leq \frac{\Gamma}{\Gamma+(n_{i}-1)}$, which can be implied by the vanilla box constraints $0\leq p_{ij}\leq 1$ plus the linear constraints $\sum_{j=1}^{n_{i}} p_{ij}=1$ and $p_{ij}-\Gamma p_{ij^{\prime}}\leq 0$ (i.e., the Rosenbaum bounds).
\end{remark}

\begin{remark}\label{rem: cross fitting}
As discussed in Section~\ref{sec: methods} of the main text, the validity of the confidence set $\Delta_{1-\alpha^{\prime}}$ requires that each balance test score $s_{ij}$ is fixed across different treatment assignments $\mathbf{Z}$ after matching. If each score $s_{ij}$ is constructed using only the measured confounders $\mathbf{X}$ and does not involve the treatment labels $\mathbf{Z}$ (e.g., when $s_{ij}$ is obtained via an unsupervised learning method such as principal component analysis), then this condition holds automatically. If each score $s_{ij}$ is constructed using both the treatment information $\mathbf{Z}$ and the measured confounders $\mathbf{X}$ (e.g., when $\widehat{s}_{ij}=\widehat{e}_{ij}=\widehat{e}(\mathbf{x}_{ij})$ or $\widehat{s}_{ij}=\widehat{p}_{ij}$, where $\widehat{p}_{ij}$ can be obtained by plugging $\widehat{e}_{ij}$ into the formulas in Remark~\ref{rem: formulas of pij}), then we can use a cross-fitting procedure to ensure that $s_{ij}$ remains invariant to different $\mathbf{Z}$ \citep{chen2023testing}. The idea is to randomly split the matched sets into two parts: we fit a propensity score model (e.g., XGBoost or random forest) on the first half and use the fitted model to construct scores $s_{ij}$ and conduct the corresponding sensitivity analysis on the second half (e.g., using Algorithm~\ref{alg: part with covariate adjustment} in the main text); meanwhile, we fit a propensity score model on the second half and use the fitted model to construct scores $s_{ij}$ and conduct the corresponding sensitivity analysis on the first half. Finally, we use a Bonferroni adjustment to combine the two sensitivity analysis results. This procedure ensures that the scores $s_{ij}$ are fixed within each sensitivity analysis, and it was adopted in our simulation studies in Section~\ref{sec: simulations} and data application in Section~\ref{sec: application}. In addition to cross-fitting, another way to ensure that $s_{ij}$ (e.g., $s_{ij}=\widehat{e}_{ij}$ or $s_{ij}=\widehat{p}_{ij}$) does not change with $\mathbf{Z}$ in the testing dataset is to fit the propensity score model using external or historical data; in this case, the fitted propensity scores do not depend on $\mathbf{Z}$ in the testing dataset, and neither do the resulting balance test scores (e.g., when $s_{ij}=\widehat{e}_{ij}$ or $s_{ij}=\widehat{p}_{ij}$).

\end{remark}

\begin{remark}\label{rem: judge the empty set}
    To judge whether $\Lambda_{\Gamma}\cap \Delta_{1-\alpha^{\prime}}=\emptyset$, we just need to solve the following convex program:
    \begin{equation*}
     \begin{split}
        \underset{\mathbf{p}}{\text{minimize}} \quad & \{S-E_{\mathbf{p}}(S)\}^{2}-\chi_{1-\alpha^{\prime}, 1}^{2}\text{var}_{\mathbf{p}}(S) \\
         \text{subject to}\quad &\sum_{j=1}^{n_{i}} p_{ij}=1, \quad\forall i,\\
        &\frac{1}{1+(n_{i}-1)\Gamma} \leq p_{ij}\leq \frac{\Gamma}{\Gamma+(n_{i}-1)}, \quad \forall i, j, \\
        &     p_{ij}-\Gamma p_{ij^{\prime}}\leq 0, \quad \forall i, j, j^{\prime},
    \end{split}
 \end{equation*}
 in which we used the fact that the constraint $\mathbf{p}\in \Lambda_{\Gamma}$ is equivalent to $\sum_{j=1}^{n_{i}} p_{ij}=1, \ \forall i$, $\frac{1}{1+(n_{i}-1)\Gamma} \leq p_{ij}\leq \frac{\Gamma}{\Gamma+(n_{i}-1)}, \ \forall i, j$, and $p_{ij}-\Gamma p_{ij^{\prime}}\leq 0, \ \forall i, j, j^{\prime}$. If the optimal value of the above convex program is greater than zero, we have $\Lambda_{\Gamma}\cap \Delta_{1-\alpha^{\prime}}=\emptyset$. Otherwise, we have $\Lambda_{\Gamma}\cap \Delta_{1-\alpha^{\prime}}\neq \emptyset$. 
\end{remark}

\begin{remark}\label{rem: equiv formulation}
In some cases, we only want to figure out whether the worst-case $p$-value under the proposed new formulation (\ref{eqn: fractional program}) is less than the significance level $\alpha$ or not (e.g., in the simulation studies). In those cases, we only need to set $c=\chi^{2}_{1-\alpha^{\prime \prime}, 1}$ in the convex program ($L_{c}$), where $\alpha^{\prime \prime}=\alpha-\alpha^{\prime}$, and then check whether the optimal value of $(L_{c})$ is greater than zero (in which case we reject the null hypothesis $H_{0}$) or not. This is because $p^{*}<\alpha \iff d^{*}>\chi^{2}_{1-\alpha^{\prime \prime}, 1}\iff$ the optimal value of the convex program ($L_{c}$) (when setting $c=\chi^{2}_{1-\alpha^{\prime \prime}, 1}$, where $\alpha^{\prime \prime}=\alpha-\alpha^{\prime}$) is greater than zero.  
\end{remark}

\begin{remark}[Connections and Differences with Existing Conditional Randomization Tests Accounting for Covariate Imbalance]
In the randomized experiments literature, there is related work on conducting conditional randomization tests that account for covariate (confounder) imbalance (e.g., \citealp{hennessy2016conditional, branson2019randomization}). For example, \citet{hennessy2016conditional} proposes a conditional randomization test that permutes treatment assignments within strata defined by categorical covariates, while \citet{branson2019randomization} develops a rejection-sampling algorithm to perform randomization tests that condition on the realized covariate balance of an experiment, as measured by statistics such as the Mahalanobis distance, allowing for general covariate types beyond the categorical case. These conditional randomization tests leverage observed covariate imbalance information to improve statistical efficiency or testing power in randomized experiments, or equivalently, to improve power of primary analyses in observational studies under the assumption of no unmeasured confounding. In contrast, our work uses covariate (confounder) imbalance information to improve power of sensitivity analysis in observational studies, where unmeasured confounding is explicitly allowed. As a result, the testing procedure proposed in our work (Algorithm~\ref{alg: part with covariate adjustment}) differs fundamentally from the permutation-based procedures of \citet{hennessy2016conditional} and \citet{branson2019randomization}. In randomized experiments, treatment assignment probabilities are known by design, which allows these methods to generate Monte Carlo draws of treatment indicators under Fisher's sharp null while conditioning on the observed covariate imbalance, either via partitioned permutations or rejection sampling, and to obtain $p$-values by directly comparing the observed test statistic to the simulated conditional permutation distribution, without requiring any multiplicity adjustment. By contrast, in sensitivity analysis for observational studies, the post-matching treatment assignment probabilities $\mathbf{p}$ are unknown and potentially biased by unmeasured confounding. Consequently, we introduce an optimization-based procedure to compute the restricted worst-case (maximal) $p$-value $p_{\Delta}^{*}$ over all $\mathbf{p}$ consistent with both the Rosenbaum bounds constraint encoded in $\Lambda_{\Gamma}$ and the post-matching covariate imbalance information encoded in $\Delta_{1-\alpha^{\prime}}$. After obtaining $p_{\Delta}^{*}$, rigorous Type-I error control requires a multiplicity adjustment that accounts for both the balance test used to construct $\Delta_{1-\alpha^{\prime}}$ and the outcome test used to obtain $p_{\Delta}^{*}$, as described in Step~3 in Section~\ref{sec: methods}.

\end{remark}

\begin{remark}\label{rem: extension}
In this work, we focus on illustrating the core idea of our approach (i.e., using the post-matching overt bias information as a valid negative control to refine the feasible set of hidden biases in a sensitivity analysis) under the classic Rosenbaum bounds sensitivity analysis framework, one of the most important sensitivity analysis frameworks in matched observational studies (or even in the whole domain of causal inference). Specifically, we showed how considering the refined feasible set $\Lambda_{\Gamma}\cap \Delta_{1-\alpha^{\prime}}$ can produce more informative sensitivity analysis results than the conventional feasible set $\Lambda_{\Gamma}$ in Rosenbaum-type sensitivity analysis. Because the confidence set $\Delta_{1-\alpha^{\prime}}$ is model-free and independent of which sensitivity analysis framework we are using, our approach can be directly applied to many other sensitivity analysis frameworks in matched observational studies. For example, \citet{fogarty2019extended} proposed a new sensitivity analysis framework for pair-matched observational studies, which extends the conventional one-parameter Rosenbaum-type sensitivity analysis by considering two sensitivity parameters: one for bounding the maximal (worst-case) biased post-matching treatment probabilities $p_{ij}$ (i.e., equivalent to the sensitivity parameter $\Gamma$ in the Rosenbaum bounds constraint) and one for bounding the expectation (i.e., average bias) of $p_{ij}$ (denoted as $\overline{\Gamma}\geq 1$). \citet{fogarty2019extended} showed that, under their proposed two-parameter sensitivity analysis framework, determining whether the worst-case (maximal) $p$-value is below the significance level or not can be formulated as solving a convex program. To apply our proposed approach to the extended sensitivity analysis framework in \citet{fogarty2019extended}, we just need to add an additional convex constraint $\mathbf{p}\in \Delta_{1-\alpha^{\prime}}$ to their formulated convex program (possibly after adjusting for multiplicity when setting the significance level for the outcome test, as discussed in Step 3 in Section~\ref{sec: methods}).

\end{remark}

\begin{remark}\label{rem: negative control general}
The negative control outcome approach of \citet{rosenbaum2023sensitivity} is formulated within the Rosenbaum bounds sensitivity analysis framework, in which the sensitivity parameter $\Gamma$ governs the extent of departure from random assignment within matched sets. While the original motivation of this approach assumes that overt bias has been sufficiently addressed by matching, the Rosenbaum bounds framework itself does not require exact post-matching confounder balance. As a result, the parameter $\Gamma$ may be interpreted more generally as capturing the aggregate effect of both hidden bias and residual post-matching overt bias. Under this interpretation, systematic associations between treatment assignment and the negative control outcome can arise not only from unmeasured confounding but also from remaining imbalance in measured confounders after matching. Therefore, the negative control outcome approach can be viewed as a diagnostic tool for detecting overall departures from effective randomization (whether due to hidden or overt bias) and for incorporating this information into a more informative sensitivity analysis.
\end{remark}

\begin{remark}
The mathematical form of the optimization formulation ($L_{c}$) is similar to that in \citet{rosenbaum2023sensitivity}, but there are two major differences. First, as mentioned earlier, the scores $s_{ij}$ in our formulation are constructed using post-matching overt bias information, whereas the corresponding scores in \citet{rosenbaum2023sensitivity} are constructed from negative control outcomes. Second, as described in Algorithm~\ref{alg: part with covariate adjustment}, the parameter $c$ in ($L_{c}$) is iteratively updated until convergence to the worst-case (minimal) squared deviate $d^{*}$ from the original formulation~(\ref{eqn: fractional program}), which in turn yields the worst-case (maximal) $p$-value $p^{*}$. In contrast, the parameter $c$ in \citet{rosenbaum2023sensitivity} is fixed at the constant $\chi^{2}_{1-\alpha^{\prime\prime}, 1}$ (with multiplicity adjustment for the negative control outcome) or $\chi^{2}_{1-\alpha, 1}$ (without multiplicity adjustment), so researchers can only determine whether $H_{0}$ is rejected in a sensitivity analysis at the nominal level $\alpha$, rather than obtain the actual $p$-value.

However, because the optimization formulation in \citet{rosenbaum2023sensitivity} is also convex when $\chi^{2}_{1-\alpha^{\prime\prime}, 1}$ or $\chi^{2}_{1-\alpha, 1}$ is replaced by any $c>0$, the proposed iterative convex programming procedure in Algorithm~\ref{alg: part with covariate adjustment} can be directly applied to their formulation to recover the actual $p$-value for testing $H_{0}$ in a sensitivity analysis informed by negative control outcomes. Specifically, it suffices to replace the scores $s_{ij}$ in Algorithm~\ref{alg: part with covariate adjustment}, which in our work are constructed from post-matching overt bias information, with scores constructed from negative control outcomes considered in \citet{rosenbaum2023sensitivity}.

\end{remark}

\begin{remark}[Confounder Balance Tables in the Simulation Studies and Data Application]

Table~\ref{tab: std} reports, for each of the five measured confounders considered in the simulation studies in Section~\ref{sec: simulations}, the average absolute standardized difference in means between the treatment and control groups across the 1000 matched datasets. Table~\ref{tab: realstd} reports, for each of the 20 measured confounders in the data application in Section~\ref{sec: application}, the absolute standardized difference in means between the treatment and control groups after matching.

\begin{table}[ht]
\small
\centering
\caption{The average post-matching absolute standardized differences in means of the five measured confounders between the treatment and control groups.}
\begin{tabular}{ccccccc}
\toprule
\multirow{2}{*}{}&\multicolumn{3}{c}{$N=500$}&\multicolumn{3}{c}{$N=1000$} \\
\cmidrule(rl){2-4} \cmidrule(rl){5-7} 
 & {Model 1} & {Model 2} & {Model 3} & {Model 1} & {Model 2} & {Model 3} \\
\midrule
$x_{1}$ & 0.088 & 0.110 & 0.099 & 0.071 & 0.100 & 0.085\\
$x_{2}$ & 0.099 & 0.126 & 0.114 & 0.083 & 0.121 & 0.104\\
$x_{3}$ & 0.043 & 0.047 & 0.046 & 0.028 & 0.032 & 0.031\\
$x_{4}$ & 0.048 & 0.054 & 0.052 & 0.033 & 0.038 & 0.036\\
$x_{5}$ & 0.109 & 0.131 & 0.121 & 0.102 & 0.132 & 0.119\\
\bottomrule
\end{tabular}
\label{tab: std}
\end{table}
\end{remark}

\begin{table}[ht]
\centering
\small
\caption{The absolute standardized differences in means of the measured confounders between the treatment and control groups, before and after matching.}
\begin{tabular}{ccc}
\toprule
\multirow{2}{*}{} &\multicolumn{2}{c}{Absolute Standardized Difference in Means} \\
\cmidrule(rl){2-3}
Measured Confounder & Before Matching & After Matching \\
\midrule
Gender  & 0.02  & 0.02 \\
Black & 0.16  & 0.05 \\
Hispanic Ethnicity & 0.12  & 0.01 \\
Baseline Test Scores & 0.48  & 0.20 \\
Dad's Education: Vocational School &  0.10 &  0.04 \\
Dad's Education: Some College &  0.10 &  0.01 \\
Dad's Education: College & 0.31 & 0.13 \\
Missingness Indicator for Dad's Education &  0.03 &  0.00 \\
Mom's Education: Vocational School &  0.03 &  0.03 \\
Mom's Education: Some College &  0.00 & 0.02 \\
Mom's Education: College & 0.26 & 0.12 \\
Missingness Indicator for Mom's Education &  0.12 & 0.02 \\
Family Income & 0.20 & 0.07 \\
Missingness Indicator for Family Income & 0.04 & 0.08 \\
Home Ownership & 0.06 &  0.00 \\
Percentage of White Students at High School & 0.12 &  0.02 \\
Urban or Rural & 0.14 & 0.04 \\
Region: Midwest & 0.15 &  0.03 \\
Region: South &  0.12 &  0.06 \\
Region: West &  0.42 &  0.04 \\
\bottomrule
\end{tabular}
\label{tab: realstd}
\end{table}

\begin{remark}
As discussed in the main text, more informative $p$-values in sensitivity analyses should occur when the worst-case allocation of post-matching treatment assignment probabilities under the conventional formulation (denoted as $\mathbf{p}_{*}=(p_{*, 11},\dots, p_{*,In_{I}})$):
\begin{equation*}
        \text{minimize}_{\mathbf{p}} \ \{T-E_{\mathbf{p}}(T)\}^{2}/\text{var}_{\mathbf{p}}(T) \quad \text{subject to } \mathbf{p}\in \Lambda_{\Gamma},
 \end{equation*}
are different from those under the proposed new formulation (denoted as $\mathbf{p}^{*}=(p^{*}_{11},\dots, p^{*}_{In_{I}})$):
\begin{equation*}
        \text{minimize}_{\mathbf{p}} \ \{T-E_{\mathbf{p}}(T)\}^{2}/\text{var}_{\mathbf{p}}(T) \quad \text{subject to } \mathbf{p}\in \Lambda_{\Gamma}\cap \Delta_{1-\alpha^{\prime}}.
 \end{equation*}
In our data application, the sample correlation between $\mathbf{p}^{*}$ and $\mathbf{p_{*}}$ is about 0.84 under $\Gamma=4$. For example, in Table~\ref{tab: p_diff}, we report the $(p_{*, i1},\dots, p_{*, i n_{i}})$ and $(p^{*}_{i1},\dots, p^{*}_{in_{i}})$ for one of the matched sets in the data application under $\Gamma=4$. Both the sample correlation $0.84$ and the results in Table~\ref{tab: p_diff} suggest an evident difference between $\mathbf{p}^{*}$ and $\mathbf{p_{*}}$, which partially explains the evident performance differences between the proposed new approach and the conventional approach in the data application.

\begin{table}[H]
\centering
\caption{An illustration of the differences in worst-case allocations of post-matching treatment assignment probabilities between the conventional formulation and the proposed new formulation, using one of the matched sets in the data application as an example ($\Gamma=4$).}
\begin{tabular}{cccc}
\toprule
 & Unit 1 & Unit 2 & Unit 3\\
\midrule
Conventional  & 0.444 & 0.111 & 0.444 \\
Proposed   & 0.305 & 0.556 & 0.139\\
\bottomrule
\end{tabular}
\label{tab: p_diff}
\end{table}
\end{remark}

\begin{remark}[Truncation of the reported $p$-value]
For notational simplicity, the main text reports $p^{*}=2\{1-\Phi(\sqrt{d^{*}})\}+\alpha^{\prime}$, which may exceed one. To ensure that the reported value lies in $[0,1]$,
we may instead use $p_{\mathrm{tr}}^{*}
=
\min\big\{1,\,
2\{1-\Phi(\sqrt{d^{*}})\}+\alpha^{\prime}\big\}$. For any significance level $\alpha\in(0,1)$, we have $p_{\mathrm{tr}}^{*}\leq\alpha$ if and only if $p^{*}\leq\alpha$. Therefore, truncation does not change the hypothesis testing decision, the corresponding Type-I error guarantees, or the simulated power. We therefore omit the truncation in the main text to simplify notation.
\end{remark}

\section*{Appendix C: Running Time of the Proposed Approach}

\textbf{Computing Environment:} All simulation experiments in Appendix C were run on a 64-bit Windows workstation equipped with a 13th Gen Intel\textsuperscript{\textregistered}\ Core\textsuperscript{\texttrademark}\ i9-13900H CPU (2.60 GHz) and 32 GB RAM. All code was implemented in \textsf{R} 4.3.1, and the optimization problems were solved using \textsf{Gurobi} Optimizer 10.0.2.

\subsection*{C.1: Comparison of the Running Times of the Conventional and Proposed Sensitivity Analysis Approaches}

In Appendix C.1, we compare the running times of the proposed sensitivity analysis approach with those of conventional sensitivity analysis approaches. Specifically, we report the running times (in seconds) of the following procedures, evaluated using the simulation settings described in Section~\ref{sec: simulations}:
\begin{itemize}
    \item \textbf{Conventional (Approximate):} The conventional sensitivity analysis approach based on the asymptotic separability algorithm of \citet{gastwirth2000asymptotic}. This method uses an analytically explicit algorithm to obtain an approximate solution to the worst-case $p$-value under the Rosenbaum bounds constraint, but necessarily ignores post-matching overt bias information.

    \item \textbf{Conventional:} The sensitivity analysis approach based on directly solving the conventional optimization formulation for determining whether the worst-case (minimal) squared deviate under the Rosenbaum bounds constraint $\mathbf{p}\in \Lambda_{\Gamma}$ exceeds the rejection threshold at nominal level $\alpha$. Relative to the approximate approach above, this optimization-based procedure yields more precise results at the cost of modestly increased computation time. Concretely, this formulation removes the overt bias constraint $\Delta_{1-\alpha^{\prime}}$ from the convex program~(\ref{eqn: convex program}) and sets the parameter $c$ to the corresponding rejection threshold (see Remark~\ref{rem: equiv formulation}), then determines whether the resulting optimal value is positive. For the purpose of computing simulated power, each simulation run only requires determining whether $H_{0}$ can be rejected, rather than computing the actual $p$-value. Thus, the reported running time approximately corresponds to that of a single iteration of Algorithm~\ref{alg: part with covariate adjustment} with $c$ fixed at the rejection threshold and without incorporating $\Delta_{1-\alpha^{\prime}}$.

    \item \textbf{Proposed (XGBoost) and Proposed (Random Forest):} The proposed sensitivity analysis approach is based on solving the new optimization formulation under both the Rosenbaum bounds constraint and the overt bias constraint, i.e., $\mathbf{p}\in \Lambda_{\Gamma}\cap \Delta_{1-\alpha^{\prime}}$. As with the conventional optimization-based approach, this procedure determines whether the optimal value of the convex program~(\ref{eqn: convex program}), with $c$ set to the rejection threshold, is positive. Again, because each simulation run only requires a rejection decision rather than the actual $p$-value, the reported running time approximately corresponds to a single iteration of Algorithm~\ref{alg: part with covariate adjustment}, now incorporating the overt bias information $\Delta_{1-\alpha^{\prime}}$. In the simulation studies, we consider two constructions of the balance test scores $s_{ij}$ used in $\Delta_{1-\alpha^{\prime}}$, one based on the XGBoost algorithm and the other based on the random forest algorithm.
\end{itemize}

\begin{table}[ht]
\centering
\footnotesize
\caption{Simulated power and average running times (in seconds; shown in parentheses) for Rosenbaum-type sensitivity analyses using the conventional and proposed approaches. For the conventional approach, we report results from both the approximate approach based on the asymptotic separability algorithm of \citet{gastwirth2000asymptotic} and the optimization-based approach.}
\resizebox{\textwidth}{!}{%
\begin{tabular}{cccccccc}
\toprule
& \multirow{2}{*}{Sample Size $N=500$}&\multicolumn{2}{c}{Model 1} &\multicolumn{2}{c}{Model 2} &\multicolumn{2}{c}{Model 3} \\ 
\cmidrule(rl){3-4} \cmidrule(rl){5-6} \cmidrule(rl){7-8}
 & & Scenario 1 & Scenario 2 &  Scenario 1 & Scenario 2 & Scenario 1 & Scenario 2 \\
\midrule
\multirow{4}{*}{$\Gamma=7$} & Conventional (Approximate) & 0.463 (0.028)  & 0.652 (0.023)  & 0.591 (0.025)  & 0.864 (0.026)  & 0.517 (0.027)  & 0.787 (0.027) \\
& Conventional & 0.453 (0.813)  & 0.643 (1.444)  & 0.578 (0.944)  & 0.857 (1.650)  & 0.499 (0.918)  & 0.774 (1.424) \\
& Proposed (XGBoost) & 0.497 (0.838)  & 0.691 (0.942)  & 0.821 (0.814)  & 0.938 (1.070)  & 0.634 (0.994)  & 0.846 (1.016) \\
& Proposed (Random Forest) & 0.516 (0.429)  & 0.716 (0.446)  & 0.821 (1.249)  & 0.945 (0.595)  & 0.649 (0.561)  & 0.852 (0.582) \\
\midrule
\multirow{4}{*}{$\Gamma=8$} & Conventional (Approximate) & 0.223 (0.029)  & 0.368 (0.028)  & 0.338 (0.027)  & 0.683 (0.027)  & 0.275 (0.028)  & 0.529 (0.028) \\
& Conventional & 0.213 (0.659)  & 0.361 (1.401)  & 0.320 (1.462)  & 0.671 (1.487)  & 0.264 (1.380)  & 0.516 (1.417) \\
& Proposed (XGBoost) & 0.239 (0.731)  & 0.397 (1.549)  & 0.578 (0.910)  & 0.813 (1.862)  & 0.346 (0.870)  & 0.590 (1.815) \\
& Proposed (Random Forest) & 0.249 (0.480)  & 0.408 (0.900)  & 0.562 (0.553)  & 0.812 (1.013)  & 0.375 (0.524)  & 0.624 (0.811) \\
\midrule
& \multirow{2}{*}{Sample Size $N=1000$}&\multicolumn{2}{c}{Model 1} &\multicolumn{2}{c}{Model 2} &\multicolumn{2}{c}{Model 3} \\ 
\cmidrule(rl){3-4} \cmidrule(rl){5-6} \cmidrule(rl){7-8}
 & & Scenario 1 & Scenario 2 &  Scenario 1 & Scenario 2 & Scenario 1 & Scenario 2 \\
\midrule
\multirow{4}{*}{$\Gamma=9$} & Conventional (Approximate) & 0.606 (0.053)  & 0.849 (0.054)  & 0.803 (0.047)  & 0.975 (0.048)  & 0.678 (0.052)  & 0.934 (0.055) \\
& Conventional & 0.594 (1.085)  & 0.842 (1.074)  & 0.794 (1.488)  & 0.975 (1.430)  & 0.666 (1.330)  & 0.928 (1.244) \\
& Proposed (XGBoost) & 0.708 (1.420)  & 0.904 (1.437)  & 0.985 (1.892)  & 0.998 (1.828)  & 0.882 (1.728)  & 0.982 (1.528) \\
& Proposed (Random Forest) & 0.770 (1.114)  & 0.932 (1.373)  & 0.991 (1.430)  & 1.000 (1.587)  & 0.929 (1.441)  & 0.990 (1.440) \\
\midrule
\multirow{4}{*}{$\Gamma=10$} & Conventional (Approximate) & 0.370 (0.052)  & 0.653 (0.053)  & 0.567 (0.047)  & 0.915 (0.048)  & 0.454 (0.049)  & 0.807 (0.055) \\
& Conventional & 0.364 (1.208)  & 0.647 (1.200)  & 0.553 (1.440)  & 0.907 (1.421)  & 0.436 (1.282)  & 0.799 (1.129) \\
& Proposed (XGBoost) & 0.458 (1.633)  & 0.725 (1.686)  & 0.928 (1.875)  & 0.989 (1.813)  & 0.688 (1.713)  & 0.909 (1.751) \\
& Proposed (Random Forest) & 0.525 (1.336)  & 0.775 (1.345)  & 0.939 (1.439)  & 0.995 (1.379)  & 0.734 (1.397)  & 0.940 (1.456) \\
\midrule
\multirow{4}{*}{$\Gamma=11$} & Conventional (Approximate) & 0.187 (0.054)  & 0.421 (0.053)  & 0.348 (0.049)  & 0.764 (0.050)  & 0.260 (0.049)  & 0.605 (0.055) \\
& Conventional & 0.182 (1.317)  & 0.408 (1.284)  & 0.335 (1.440)  & 0.752 (1.418)  & 0.254 (1.192)  & 0.591 (1.132) \\
& Proposed (XGBoost) & 0.242 (1.560)  & 0.507 (1.700)  & 0.785 (1.875)  & 0.954 (1.799)  & 0.459 (1.720)  & 0.767 (1.735) \\
& Proposed (Random Forest) & 0.288 (1.346)  & 0.559 (1.344)  & 0.822 (1.535)  & 0.967 (1.493)  & 0.506 (1.416)  & 0.814 (1.424) \\
\bottomrule
\end{tabular}
}
\label{tab: more simulated running time}

\end{table}

Table~\ref{tab: more simulated running time} reports simulated power and average running times (in seconds; shown in parentheses) for the Rosenbaum-type sensitivity analysis procedures described above, across the simulation settings considered in Section~\ref{sec: simulations} of the main text. The approximate conventional approach based on the asymptotic separability algorithm of \citet{gastwirth2000asymptotic} is consistently the fastest, reflecting its analytically explicit nature and the absence of numerical optimization. The optimization-based conventional approach, which directly solves the corresponding convex program under the Rosenbaum bounds constraint, incurs modest additional computational cost but remains below two seconds per simulation run in most settings. The proposed sensitivity analysis approaches, which additionally incorporate the overt bias constraint $\Delta_{1-\alpha^{\prime}}$, require training the balance test scores $s_{ij}$ used in $\Delta_{1-\alpha^{\prime}}$ (via either the XGBoost or random forest algorithm) and solving the resulting constrained convex program, and therefore often exhibit longer running times than both conventional approaches. Nevertheless, these running times are stable across simulation scenarios and are typically on the order of one to two seconds per simulation run, indicating that the added computational burden is moderate for determining whether to reject $H_{0}$ (roughly speaking, corresponding to a single iteration of Algorithm~\ref{alg: part with covariate adjustment}). Finally, although not the focus here, the optimization-based and approximate conventional approaches yield very similar simulated power across the considered settings, confirming the accuracy of the asymptotic separability algorithm of \citet{gastwirth2000asymptotic}. Because the optimization-based approach provides more precise solutions, simulated power results reported in Section~\ref{sec: simulations} of the main text are based on the optimization-based approach.

To place the running times of the sensitivity analysis procedures in context, Table~\ref{tab: matching running time} reports the average running times of the matching procedure used to construct the matched datasets in Section~\ref{sec: simulations}, based on the optimal full matching algorithm \citep{hansen2004full,hansen2006optimal}. Across all simulation scenarios, the matching procedure is computationally efficient, with average running times below one second for $N=500$ and on the order of one second for $N=1000$. Comparing these results with the running times reported in Table~\ref{tab: more simulated running time}, we find that a single iteration of the proposed optimization-based sensitivity analysis has a computational cost that is comparable to, and in some settings slightly larger than, that of the matching procedure itself. This comparison indicates that the additional computational burden introduced by each iteration of the proposed approach is roughly of the same order of magnitude as the cost of matching.

\begin{table}[H]
\small
\centering
\caption{Average running times (in seconds) of the matching procedures in the simulation settings in Section~\ref{sec: simulations}, based on the optimal full matching algorithm \citep{hansen2004full, hansen2006optimal}. }
\begin{tabular}{ccccccc}
\toprule
 &\multicolumn{2}{c}{Model 1} &\multicolumn{2}{c}{Model 2} &\multicolumn{2}{c}{Model 3} \\ 
\cmidrule(rl){2-3} \cmidrule(rl){4-5} \cmidrule(rl){6-7}
Sample Size & Scenario 1 & Scenario 2 &  Scenario 1 & Scenario 2 & Scenario 1 & Scenario 2 \\
\midrule
$N=500$ & 0.333 & 0.277 & 0.315 & 0.244 & 0.274 & 0.403 \\
$N=1000$ & 1.099 & 1.234 & 0.901 & 1.005 & 1.041 & 1.084\\

\bottomrule
\end{tabular}

\label{tab: matching running time}

\end{table}

\subsection*{C.2: Running Time of the Proposed Approach in the Data Application}

In Appendix C.1, we showed that the proposed approach is computationally efficient for determining whether $H_{0}$ should be rejected (i.e., for deciding whether the $p$-value $p^{*}$ is less than the nominal level $\alpha$), because this only requires solving a single convex program in (\ref{eqn: convex program}) with $c$ fixed at the rejection threshold corresponding to level $\alpha$. In contrast, to compute the actual $p$-value $p^{*}$ (rather than only deciding whether $p^{*}<\alpha$), we apply the iterative convex programming procedure in Algorithm~\ref{alg: part with covariate adjustment}, which solves a sequence of convex programs of the form (\ref{eqn: convex program}) with an iteratively updated parameter $c$. Table~\ref{tab: runtime data analysis} reports the running times (in seconds) of Algorithm~\ref{alg: part with covariate adjustment} (from inputting the matched data to the final output) for the data application in Section~\ref{sec: application} (sample size $N=1819$). As shown in Table~\ref{tab: runtime data analysis}, the proposed iterative convex programming procedure converges in fewer than 10 iterations across the $\Gamma$ values considered, and the total running time is within 30 seconds for each $\Gamma$.

\begin{table}[ht]
\small
    \centering
    \caption{Running times (in seconds) of the proposed sensitivity analysis approach in the data application in Section~\ref{sec: application} (sample size $N=1819$).}
    \begin{tabular}{crrr}
        \toprule
        $\Gamma$ 
        & Total Running Time 
        & Number of Iterations 
        & Average Time per Iteration \\
        \midrule
        $3.50$ & 20.060 & 8 & 2.507 \\
        $3.75$ & 19.578 & 6 & 3.263 \\
        $4.00$ & 18.859 & 9 & 2.095 \\
        $4.25$ & 12.583 & 9 & 1.398 \\
        \bottomrule
    \end{tabular}
    \label{tab: runtime data analysis}
\end{table}

\section*{Appendix D: Generalization to Multivariate Balance Test Scores}

In the main text, we illustrate the principles of our proposed approach using scalar summaries $s_{ij}$ of post-matching confounder imbalance, such as setting $s_{ij}=\widehat{p}_{ij}$, where $\widehat{p}_{ij}$ denotes the estimated post-matching treatment assignment probability. These quantities $s_{ij}$ are referred to as balance test scores because they are used to construct the confidence set $\Delta_{1-\alpha^{\prime}}$ for the true post-matching treatment assignment probabilities $\mathbf{p}=(p_{11},\dots, p_{In_{I}})$. Importantly, the proposed framework can be readily generalized to accommodate multivariate measures of post-matching confounder imbalance. In this more general setting, each unit $ij$ is associated with multiple balance test scores $s_{ij1},\dots, s_{ijK}$, where $K$ may exceed one. For example, when there are exactly $K$ measured confounders, one natural choice is to set $s_{ijk}=x_{ijk}$, where $x_{ijk}$ denotes the $k$-th measured confounder for unit $j$ in matched set $i$. Another important setting arises when balance test scores are constructed using multiple propensity score learners to estimate $\widehat{p}_{ij}$. For instance, suppose that researchers consider $K$ candidate propensity score learning algorithms, such as a logistic regression model, XGBoost, random forest, or neural network. Rather than selecting a single algorithm to construct a univariate balance test score $s_{ij}=\widehat{p}_{ij}$ as in the main text, researchers may wish to incorporate information from multiple learners simultaneously by defining multivariate balance test scores $s_{ij1},\dots, s_{ijK}$, where $s_{ijk}=\widehat{p}_{ij}^{(k)}$ and $\widehat{p}_{ij}^{(k)}$ denotes the estimate of $p_{ij}$ obtained from the $k$-th propensity score learning algorithm.

To formally establish the theoretical foundation for this extension, suppose that each unit $ij$ is associated with $K$ balance test scores $s_{ij1},\dots, s_{ijK}$, where $K\geq 2$. Correspondingly, let $S_{k}=\sum_{i=1}^{I}\sum_{j=1}^{n_{i}}Z_{ij}s_{ijk}$ denote the $k$-th balance test statistic based on the $k$-th balance test score, for $k=1,\dots, K$. Let $\alpha^{\prime}\in (0,\alpha)$ be a prespecified total significance level allocated across the $K$ balance tests $S_{1}, \dots, S_{K}$. For each balance test $S_{k}$, we define $\Delta_{1-\alpha^{\prime}/K}^{(k)}=\big\{\mathbf{p}^{\prime}=(p^{\prime}_{11}, \dots, p^{\prime}_{In_{I}}) \in [0,1]^{N}: \sum_{j=1}^{n_{i}}p^{\prime}_{ij}=1 \text{ for all $i$, } \{S_{k}-E_{\mathbf{p}^{\prime}}(S_{k})\}^{2}-\chi_{1-\alpha^{\prime}/K, 1}^{2}\text{var}_{\mathbf{p}^{\prime}}(S_{k})\leq 0 \big\}$, where $E_{\mathbf{p}^{\prime}}(S_{k})=\sum_{i\in \mathcal{I}_{1}}\sum_{j=1}^{n_{i}}p^{\prime}_{ij}s_{ijk}+\sum_{i\in \mathcal{I}_{2}}\sum_{j=1}^{n_{i}}(1-p^{\prime}_{ij})s_{ijk}$ and $\text{var}_{\mathbf{p}^{\prime}}(S_{k})=\sum_{i=1}^{I}\big\{ \sum_{j=1}^{n_i}p^{\prime}_{ij}s_{ijk}^2-(\sum_{j=1}^{n_i}p^{\prime}_{ij}s_{ijk})^2\big \}$. Thus, $\Delta_{1-\alpha^{\prime}/K}^{(k)}$ represents a $100(1-\alpha^{\prime}/K)\%$ confidence set for $\mathbf{p}$ constructed using the $k$-th balance test statistic $S_{k}$. The adjusted level $\alpha^{\prime}/K$ reflects the Bonferroni adjustment applied to account for multiplicity across the $K$ balance tests. We then consider the following quadratic fractional program:
\begin{equation}\label{eqn: multivariate fractional program}
        \text{minimize}_{\mathbf{p}} \ \{T-E_{\mathbf{p}}(T)\}^{2}/\text{var}_{\mathbf{p}}(T) \quad \text{subject to } \mathbf{p}\in \Lambda_{\Gamma}\cap \left(\cap_{k=1}^{K}\Delta_{1-\alpha^{\prime}/K}^{(k)}\right).
 \end{equation} 
Let $d^{\diamond}$ denote the optimal value of the quadratic fractional program in (\ref{eqn: multivariate fractional program}), and define the corresponding $p$-value as $p^{\diamond}=2\{1-\Phi(\sqrt{d^{\diamond}})\}+\alpha^{\prime}$. By the same arguments used in Lemma~\ref{thm: validity of A}, each $\Delta_{1-\alpha^{\prime}/K}^{(k)}$ is a valid $100(1-\alpha^{\prime}/K)\%$ confidence set for $\mathbf{p}$. Consequently, the intersection $\cap_{k=1}^{K}\Delta_{1-\alpha^{\prime}/K}^{(k)}$ forms a valid $100(1-\alpha^{\prime})\%$ confidence set for $\mathbf{p}$. Following the same reasoning as in Theorem~\ref{thm: validity}, with $\Delta_{1-\alpha^{\prime}}$ replaced by $\cap_{k=1}^{K}\Delta_{1-\alpha^{\prime}/K}^{(k)}$, the resulting $p$-value $p^{\diamond}$ is valid under the Rosenbaum bounds constraint.

Moreover, Algorithm~\ref{alg: part with covariate adjustment} can be directly adapted to solve the quadratic fractional program in (\ref{eqn: multivariate fractional program}). Specifically, for any $c\geq 0$, we consider
\begin{equation}\label{eqn: multivariate convex program}
        \text{minimize}_{\mathbf{p}} \ \{T-E_{\mathbf{p}}(T)\}^{2}-c\times \text{var}_{\mathbf{p}}(T) \quad \text{subject to } \mathbf{p}\in \Lambda_{\Gamma}\cap \left(\cap_{k=1}^{K}\Delta_{1-\alpha^{\prime}/K}^{(k)}\right),
 \end{equation}
which can be equivalently written as the following quadratically constrained quadratic program:
\begin{equation*}
     \begin{split}
        \underset{\mathbf{p}}{\text{minimize}} \quad & \{T-E_{\mathbf{p}}(T)\}^{2}-c\times \text{var}_{\mathbf{p}}(T) \quad \quad (L^{\diamond}_{c}) \\
         \text{subject to}\quad &\sum_{j=1}^{n_{i}} p_{ij}=1, \quad\forall i,\\
        &0 \leq p_{ij}\leq 1, \quad \forall i, j,\\
        & p_{ij}-\Gamma p_{ij^{\prime}}\leq 0, \quad \forall i, j, j^{\prime},\\
        & \{S_{k}-E_{\mathbf{p}}(S_{k})\}^{2}-\chi_{1-\alpha^{\prime}/K, 1}^{2} \text{var}_{\mathbf{p}}(S_{k}) \leq 0, \quad \forall k\in \{1,\dots, K\}.
     \end{split}
 \end{equation*}
Because each $\Delta_{1-\alpha^{\prime}/K}^{(k)}$ is convex, by the same argument as in the proof of Theorem~\ref{thm: convexity}, their intersection $\cap_{k=1}^{K}\Delta_{1-\alpha^{\prime}/K}^{(k)}$ is also convex. Consequently, the quadratically constrained quadratic program in (\ref{eqn: multivariate convex program}) is convex. Following the same reasoning as in the proof of Theorem~\ref{thm: convergence}, replacing the convex program (\ref{eqn: convex program}) with (\ref{eqn: multivariate convex program}) in Algorithm~\ref{alg: part with covariate adjustment} yields a valid procedure for computing the $p$-value $p^{\diamond}$ when multivariate balance test scores $s_{ij1},\dots, s_{ijK}$ are used.

Table~\ref{tab: vector_simulated} reports preliminary simulation results extending the simulation studies in Section~\ref{sec: simulations} from univariate to multivariate balance test scores, with sample size $N=500$. Recall that in Section~\ref{sec: simulations}, we use two widely adopted machine learning algorithms, XGBoost and random forest, to construct the balance test scores $s_{ij}=\widehat{p}_{ij}$ based on post-matching confounder imbalance. Beyond selecting one of these algorithms, our framework allows both to be incorporated simultaneously by defining vector-valued balance test scores $\mathbf{s}_{ij}=(s_{ij1}, s_{ij2})$, where $s_{ij1}$ is constructed using XGBoost and $s_{ij2}$ using random forest. Simulated power of the proposed approach based on such multivariate balance test scores is reported in Table~\ref{tab: vector_simulated}. The results show that, compared with the conventional approach, the proposed approach incorporating multivariate balance test scores continues to offer improvements in statistical power across many settings.

\begin{table}[ht]
\centering
\footnotesize
\caption{Simulated power of the two approaches to Rosenbaum-type sensitivity analysis: the conventional approach and the proposed approach with $\mathbf{s}_{ij}=(s_{ij1}, s_{ij2})$.}
\resizebox{\textwidth}{!}{\begin{tabular}{cccccccc}
\toprule
& \multirow{2}{*}{}&\multicolumn{2}{c}{Model 1} &\multicolumn{2}{c}{Model 2} &\multicolumn{2}{c}{Model 3} \\ 
\cmidrule(rl){3-4} \cmidrule(rl){5-6} \cmidrule(rl){7-8}
 & & Scenario 1 & Scenario 2 &  Scenario 1 & Scenario 2 & Scenario 1 & Scenario 2 \\
\midrule
\multirow{2}{*}{$\Gamma=7$} & Conventional & 0.453 & 0.643 & 0.578 & 0.857 & 0.499 & 0.774\\
& Proposed (Multivariate) & 0.499 & 0.694 & 0.812 & 0.939 & 0.624 & 0.843\\
\midrule
\multirow{2}{*}{$\Gamma=8$} & Conventional & 0.213 & 0.361 & 0.320 & 0.671 & 0.264 & 0.516\\
& Proposed (Multivariate) & 0.243 & 0.395 & 0.551 & 0.803 & 0.355 & 0.591\\
\bottomrule
\end{tabular}}
\label{tab: vector_simulated}

\end{table}

After demonstrating the feasibility and validity of incorporating multivariate balance test scores $\mathbf{s}_{ij}=(s_{ij1},\dots, s_{ijK})$ through both theoretical and numerical investigations, a natural next question concerns how best to exploit this added flexibility to further improve statistical power. In particular, relative to the univariate case, incorporating multivariate balance test scores incurs a multiplicity adjustment cost, as the significance level allocated to each component confidence set is reduced from $\alpha^{\prime}$ to $\alpha^{\prime}/K$. This tradeoff motivates several important directions for future research, including developing principled guidance on when to prefer univariate versus multivariate balance test scores, and how to select or construct the components of a multivariate balance test score $\mathbf{s}_{ij}$. For example, when $\mathbf{s}_{ij}$ is defined using a subset of measured confounders $\mathbf{x}_{ij}$, it is essential to develop a systematic and powerful procedure for identifying informative subsets of $\mathbf{x}_{ij}$, particularly in settings with a large number of measured confounders.

\section*{Appendix E: Preliminary Investigations of Alternative Balance Test Scores Incorporating Additional Knowledge on Potential Outcomes}

As demonstrated by the simulation studies and data application in the main text, setting the balance test score to $s_{ij}=\widehat{p}_{ij}$ (i.e., the estimated post-matching treatment assignment probability based on overt bias information) can substantially improve power of sensitivity analysis across a wide range of scenarios. This choice is also intuitively appealing. The primary motivation for introducing balance test scores $s_{ij}$ is to leverage observed covariate imbalance to obtain a more informative feasible set for $\mathbf{p}=(p_{11},\dots,p_{In_{I}})$ than that implied by the Rosenbaum bounds alone. From this perspective, directly setting $s_{ij}=\widehat{p}_{ij}$ provides a natural and principled way to incorporate available prior information about treatment assignment probabilities into the sensitivity analysis framework.

In practice, beyond setting $s_{ij}=\widehat{p}_{ij}$, there may exist other meaningful choices of balance test scores $s_{ij}$, particularly when prior knowledge about potential outcomes is available. In this section, we conduct preliminary investigations of an alternative choice of $s_{ij}$ that incorporates information related to potential outcomes. It is well known that incorporating information about potential outcomes contributed by measured confounders (observed covariates) into outcome test scores $q_{ij}$ can substantially improve statistical power in many settings; such procedures are commonly referred to as covariate, covariance, or regression adjustment for outcome tests \citep{rosenbaum2002covariance, fogarty2019biased, li2020rerandomization}. Specifically, when testing Fisher's sharp null hypothesis $H_{0}$, researchers use the measured confounders $\mathbf{X}$ to fit a prediction model for $Y_{ij}(0)$ (noting that $Y_{ij}=Y_{ij}(0)$ under $H_{0}$) and obtain fitted outcomes $\widehat{Y}_{ij}(0)$. A prediction function for $Y_{ij}(0)$ based on confounders is often referred to as a \textit{prognostic score} \citep{hansen2008prognostic}; consequently, $\widehat{Y}_{ij}(0)$ can be viewed as an estimated prognostic score. Instead of using the original outcomes $Y_{ij}$ to construct outcome test scores $q_{ij}$, researchers may use the residuals
$\widehat{\epsilon}_{ij}=Y_{ij}-\widehat{Y}_{ij}(0)$.

In parallel, an interesting question is whether incorporating overt bias information through estimated prognostic scores $\widehat{Y}_{ij}(0)$, rather than through estimated post-matching treatment assignment probabilities (post-matching propensity scores) $\widehat{p}_{ij}$, can also substantially improve power of sensitivity analysis. To obtain empirical evidence, we extend the simulation studies in Section~\ref{sec: simulations} of the main text. Specifically, across the simulation scenarios considered in Section~\ref{sec: simulations} (with sample size $N=500$), we consider the following four settings:
\begin{itemize}
    \item \textbf{Setting 1 (Baseline):} In this baseline setting, estimated prognostic scores $\widehat{Y}_{ij}(0)$ based on $\mathbf{X}$ are not incorporated into either the balance test scores $s_{ij}$ (i.e., we still set $s_{ij}=\widehat{p}_{ij}$) or the outcome test scores $q_{ij}$ (i.e., we still set $q_{ij}=Y_{ij}$). This setting therefore coincides exactly with the setup considered in Section~\ref{sec: simulations} of the main text, and the corresponding simulation results are identical to those reported in Table~\ref{tab: more simulated power}.
    
    \item \textbf{Setting 2 (Incorporating Estimated Prognostic Scores into the Balance Test):} In this setting, estimated prognostic scores based on $\mathbf{X}$ (estimated using XGBoost and random forest) are incorporated into the balance test scores $s_{ij}$ only (i.e., setting $s_{ij}=\widehat{Y}_{ij}(0)$), but not into the outcome test scores $q_{ij}$ (i.e., we still set $q_{ij}=Y_{ij}$). Here, post-matching overt bias is characterized by imbalance in estimated prognostic scores rather than observed imbalance in estimated post-matching propensity scores as in Setting~1.
    
    \item \textbf{Setting 3 (Incorporating Oracle Prognostic Scores into the Balance Test):} In this setting, as an oracle benchmark for Setting 2, oracle prognostic scores $Y_{ij}(0)$ are incorporated into the balance test scores $s_{ij}$ (i.e., setting $s_{ij}=Y_{ij}(0)$), but not into the outcome test scores $q_{ij}$ (i.e., we still set $q_{ij}=Y_{ij}$). Here, post-matching overt bias is characterized by imbalance in oracle prognostic scores rather than observed imbalance in estimated post-matching propensity scores as in Setting~1.
    
    \item \textbf{Setting 4 (Incorporating Prognostic Scores into the Outcome Test):} In this setting, estimated prognostic scores based on $\mathbf{X}$ (estimated using XGBoost and random forest) are not incorporated into the balance test scores $s_{ij}$ (i.e., we still set $s_{ij}=\widehat{p}_{ij}$) but into the outcome test scores $q_{ij}$ (i.e., setting $q_{ij}=\widehat{\epsilon}_{ij}=Y_{ij}-\widehat{Y}_{ij}(0)$).
\end{itemize}

\begin{table}[ht]
\centering
\caption{Simulated power of the two approaches to Rosenbaum-type sensitivity analysis in the four considered settings for investigating the incorporation of prognostic scores (estimated or oracle): the conventional approach and the proposed approach.}
\label{tab:noise_simulated}
\resizebox{\textwidth}{!}{%
\begin{tabular}{cllcccccc}
\toprule
& & & \multicolumn{2}{c}{Model 1} & \multicolumn{2}{c}{Model 2} & \multicolumn{2}{c}{Model 3} \\
\cmidrule(rl){4-5} \cmidrule(rl){6-7} \cmidrule(rl){8-9}
$\Gamma$ & Setting & Algorithm & Scenario 1 & Scenario 2 & Scenario 1 & Scenario 2 & Scenario 1 & Scenario 2 \\
\midrule
\multirow{11}{*}{$\Gamma=7$} & \multirow{3}{*}{Setting 1} & Conventional & 0.453 & 0.643 & 0.578 & 0.857 & 0.499 & 0.774 \\ 
& & Proposed (XGBoost) & 0.497 & 0.691 & 0.821 & 0.938 & 0.634 & 0.846 \\ 
& & Proposed (Random Forest) & 0.516 & 0.716 & 0.821 & 0.945 & 0.649 & 0.852 \\ 
\cmidrule(l){2-9}
& \multirow{3}{*}{Setting 2} & Conventional & 0.453 & 0.643 & 0.578 & 0.857 & 0.499 & 0.774 \\ 
& & Proposed (XGBoost) & 0.453 & 0.643 & 0.581 & 0.857 & 0.499 & 0.774 \\ 
& & Proposed (Random Forest) & 0.453 & 0.643 & 0.581 & 0.857 & 0.499 & 0.774 \\ 
\cmidrule(l){2-9}
& \multirow{2}{*}{Setting 3} & Conventional & 0.453 & 0.643 & 0.578 & 0.857 & 0.499 & 0.774 \\ 
& & Proposed (Oracle $Y_{ij}(0)$) & 0.487 & 0.658 & 0.598 & 0.858 & 0.521 & 0.784 \\ 
\cmidrule(l){2-9}
& \multirow{3}{*}{Setting 4} & Conventional & 0.453 & 0.643 & 0.578 & 0.857 & 0.499 & 0.774 \\ 
& & Proposed (XGBoost) & 0.658 & 0.829 & 0.907 & 0.984 & 0.783 & 0.921 \\ 
& & Proposed (Random Forest) & 0.685 & 0.839 & 0.919 & 0.986 & 0.789 & 0.926 \\ 
\midrule
\multirow{11}{*}{$\Gamma=8$} & \multirow{3}{*}{Setting 1} & Conventional & 0.213 & 0.361 & 0.320 & 0.671 & 0.264 & 0.516 \\ 
& & Proposed (XGBoost) & 0.239 & 0.397 & 0.578 & 0.813 & 0.346 & 0.590 \\ 
& & Proposed (Random Forest) & 0.249 & 0.408 & 0.562 & 0.812 & 0.375 & 0.624 \\ 
\cmidrule(l){2-9}
& \multirow{3}{*}{Setting 2} & Conventional & 0.213 & 0.361 & 0.320 & 0.671 & 0.264 & 0.516 \\ 
& & Proposed (XGBoost) & 0.213 & 0.361 & 0.320 & 0.671 & 0.264 & 0.517 \\ 
& & Proposed (Random Forest) & 0.213 & 0.361 & 0.320 & 0.671 & 0.264 & 0.517 \\ 
\cmidrule(l){2-9}
& \multirow{2}{*}{Setting 3} & Conventional & 0.213 & 0.361 & 0.320 & 0.671 & 0.264 & 0.516 \\ 
& & Proposed (Oracle $Y_{ij}(0)$) & 0.226 & 0.365 & 0.331 & 0.674 & 0.271 & 0.520 \\ 
\cmidrule(l){2-9}
& \multirow{3}{*}{Setting 4} & Conventional & 0.213 & 0.361 & 0.320 & 0.671 & 0.264 & 0.516 \\ 
& & Proposed (XGBoost) & 0.384 & 0.574 & 0.734 & 0.898 & 0.508 & 0.741 \\ 
& & Proposed (Random Forest) & 0.401 & 0.604 & 0.727 & 0.906 & 0.524 & 0.765 \\ 
\bottomrule
\end{tabular}
} 
\end{table}

Table~\ref{tab:noise_simulated} reports simulated power of the conventional and proposed sensitivity analysis approaches across Settings~1--4 under the simulation setup of Section~\ref{sec: simulations} in the main text. The four settings exhibit clearly distinct power patterns that help clarify the respective roles of propensity scores and prognostic scores within the proposed framework. Setting~1 reproduces the main findings from Section~\ref{sec: simulations}, confirming that incorporating estimated post-matching treatment assignment probabilities $\widehat{p}_{ij}$ into the balance test scores substantially improves power relative to the conventional Rosenbaum-type sensitivity analysis. Settings~2 and 3 show at most modest power gains when propensity-score-based balance test scores are replaced by either estimated prognostic scores ($s_{ij}=\widehat{Y}_{ij}(0)$) or oracle prognostic scores ($s_{ij}=Y_{ij}(0)$). These results suggest that, in the settings considered, the additional constraints on the feasible set for $\mathbf{p}$ imposed by prognostic scores alone offer limited power benefits beyond those obtained under the Rosenbaum bounds. Setting~4, however, demonstrates substantial power improvements when estimated prognostic scores are incorporated into the outcome test via covariate adjustment (i.e., using $\widehat{\epsilon}_{ij}$ to construct $q_{ij}$). Taken together, these results highlight a clear conceptual distinction between the roles of propensity scores and prognostic scores: propensity scores are naturally suited for constructing balance test scores that sharpen constraints on post-matching treatment assignment probabilities, whereas prognostic scores are most effective when used for outcome adjustment to reduce outcome variability, and combining these two components offers a promising direction for further improving power of sensitivity analysis.

\section*{Appendix F: Performance of the Proposed Approach with External Data}

As noted in Section~\ref{sec: methods} and Remark~\ref{rem: cross fitting} in Appendix~B, the propensity score estimates used to form the balance test scores $s_{ij}$ can be obtained either by cross-fitting on the matched sample or by leveraging external data (e.g., historical datasets). When suitable external data are available, cross-fitting is not required: the propensity score model can be trained on data independent of the matched sample, so the resulting $\widehat{p}_{ij}$ (and hence $s_{ij}$) are effectively fixed with respect to the treatment assignments $\mathbf{Z}$ in the analysis sample. This independence allows us to use the full matched sample both to construct the confounder-balance-induced constraint $\Delta_{1-\alpha^{\prime}}$ and to conduct the outcome test under the Rosenbaum bounds framework, thereby improving the effective sample size. This motivates a practical question: when sufficient external data are available, should we construct $s_{ij}$ using propensity scores estimated from external data, or should we instead estimate propensity scores using the internal matched sample with cross-fitting?

To address this question, recall that in designing a trustworthy sensitivity analysis workflow (from balance test to outcome test under the Rosenbaum bounds framework), two considerations are both important:
\begin{itemize}
    \item \textbf{Consideration (i): Power of the balance test for rejecting understated values of the sensitivity parameter $\Gamma$.} The balance test conducted prior to the outcome analysis should be able to reject understated values of $\Gamma$. Otherwise, the sensitivity analysis may proceed under a value of $\Gamma$ that is too small, yielding overly optimistic conclusions and diminishing the scientific relevance of the reported results \citep{chen2023testing}.

    \item \textbf{Consideration (ii): Power of the outcome test for detecting the actual treatment effect under each prespecified value of $\Gamma$.} For each prespecified $\Gamma$, the resulting Rosenbaum-type outcome test (based on the worst-case $p$-value under the Rosenbaum bounds constraint) should retain as much power as possible to detect the true effect. Otherwise, even if the balance test supports a reasonable (non-falsified) choice of $\Gamma$, the subsequent sensitivity analysis may be overly conservative and yield uninformative $p$-values and confidence intervals \citep{rosenbaum2004design, rosenbaum2020design}.
\end{itemize}

Although external data can increase effective sample size relative to using internal data only (with cross-fitting), it need not be preferred in general. In particular, if there is distribution shift between the external and internal datasets (e.g., differences in the propensity score model $\mathbf{Z}\mid \mathbf{X},\mathbf{U}$ and/or differences in the marginal distribution of $(\mathbf{X},\mathbf{U})$), then propensity scores estimated from external data may be systematically biased relative to those that would be learned from the internal matched sample. Such misspecification can compromise consideration~(i) by weakening the balance test's ability to reject understated values of $\Gamma$, and it can also compromise consideration~(ii) by attenuating, or even negating, the power gains one might otherwise obtain by avoiding cross-fitting.

To clarify these points, we conduct a preliminary simulation study that builds on the design in Section~\ref{sec: simulations}. We consider two settings for generating the external data:
\begin{itemize}
    \item \textbf{Setting 1 (Idealized External Data):} The external and internal datasets share the same data-generating process for $(\mathbf{x}_{n},u_{n})$ and the same propensity score model, namely those described in Section~\ref{sec: simulations} of the main text.
    
    \item  \textbf{Setting 2 (External Data with Moderate Distribution Shift):} Both the data-generating process for $(\mathbf{x}_{n},u_{n})$ and the propensity score model $z_{n}\mid (\mathbf{x}_{n},u_{n})$ differ between the external and internal datasets. Specifically, we introduce external-data noise $u_{e,n}\overset{\text{i.i.d.}}{\sim} N(0,1)$ into the main terms of measured confounders in the propensity score model and also change the relative magnitudes of the main terms within the function $\phi(\mathbf{x}_{n})$ in the conditional distribution $z_{n}\mid \mathbf{x}_{n}$. Concretely, for Models 1--3 in Section~\ref{sec: simulations}, we change $\phi(\mathbf{x}_{n})$ from $\phi(\mathbf{x}_{n})=0.3 (x_{n1} +x_{n2}) +0.6 \cos x_{n3} +0.4 (|x_{n4}|+x_{n5})+|x_{n1}  x_{n2}| +x_{n1} x_{n3}+x_{n4} x_{n5}$ to $\phi(\mathbf{x}_{n}) = 0.6 (x_{n1} + u_{e,n} + x_{n2} + u_{e,n}) + 0.3 (\cos x_{n3} + u_{e,n}) + 0.4 (|x_{n4}| + u_{e,n} + x_{n5} + u_{e,n}) + |x_{n1} x_{n2}| + x_{n1} x_{n3} + x_{n4} x_{n5} + u_{e,n}$. Thus, although the external and internal datasets remain highly correlated, they are not generated from the same propensity score model, and a moderate distribution shift is present.
\end{itemize}

We consider the case where the external and internal sample sizes are the same, with $N=500$ in each dataset. For a fair comparison, we use the random forest algorithm to train the propensity score model for both the internal data and external data settings. Unlike the internal data analysis in Section~\ref{sec: simulations}, here we do not perform cross-fitting when fitting the propensity score model. Under two-fold cross-fitting, the training and analysis roles are alternated across folds so that no observations are discarded; nevertheless, cross-fitting can reduce effective sample size because fold-specific analyses require a Bonferroni adjustment across folds (Remark~\ref{rem: cross fitting} in Appendix~B). In contrast, when external data are available, we fit the propensity score model using the external dataset and then apply the fitted model to the internal matched sample to obtain $\widehat{p}_{ij}$, thereby eliminating the need to split the matched sample into two folds.

We first examine power of the outcome test under the Rosenbaum bounds framework (Consideration~(ii)), comparing the proposed approach based on internal data with cross-fitting to the proposed approach based on external data (Settings~1 and~2). Table~\ref{tab: external outcome power} reports simulated power for the proposed approach under each simulation scenario (for the analysis sample) in Section~\ref{sec: simulations}. The results suggest that when high-quality external data are available, using external data to construct $\widehat{p}_{ij}$ can further increase power of the outcome test (i.e., power of sensitivity analysis) under the proposed approach. Comparing Settings~1 and~2 for generating external data, even a moderate distribution shift attenuates the power gains from using external data, consistent with imperfect transfer of externally trained propensity score models to the internal matched sample.

\begin{table}[ht]
\centering
\caption{Simulated power of the outcome test (under the Rosenbaum bounds framework) using balance test scores constructed from internal data only (with cross-fitting) and from external data under Setting~1 (no distribution shift) and Setting~2 (moderate distribution shift). The internal and external sample sizes are equal ($N=500$).}
\label{tab: external outcome power}
\resizebox{\textwidth}{!}{%
\begin{tabular}{cccccccc}
\toprule
& & \multicolumn{2}{c}{Model 1} & \multicolumn{2}{c}{Model 2} & \multicolumn{2}{c}{Model 3} \\
\cmidrule(rl){3-4} \cmidrule(rl){5-6} \cmidrule(rl){7-8}
 & Data Setting & Scenario 1 & Scenario 2 & Scenario 1 & Scenario 2 & Scenario 1 & Scenario 2 \\
\midrule
\multirow{3}{*}{$\Gamma=2$} & Internal Data Only & 0.816 & 0.763 & 1.000 & 0.998 & 0.983 & 0.966\\
& External Data (Setting 2) & 0.570 & 0.457 & 0.999 & 0.988 & 0.945 & 0.880\\
\midrule
\multirow{3}{*}{$\Gamma=3$} & Internal Data Only & 0.192 & 0.142 & 0.933 & 0.830 & 0.634 & 0.487\\
& External Data (Setting 2) & 0.034 & 0.028 & 0.864 & 0.650 & 0.399 & 0.239\\
\bottomrule
\end{tabular}
} 
\end{table}

We next examine the ability of the balance test to reject understated values of $\Gamma$ (Consideration~(i)), such as $\Gamma=2$ or $3$. Table~\ref{tab: external balance power comparison} reports simulated power of the balance test when the balance test scores are constructed using internal data only (with cross-fitting) versus external data with moderate distribution shift (Setting~2). The results indicate that even under a moderate distribution shift, the balance test based on external-data-derived scores can be substantially less powerful than the balance test based on internal-data-derived scores. This increases the risk that understated values of $\Gamma$ are not rejected prior to conducting the outcome analysis.

\begin{table}[ht]
\centering
\caption{Simulated power of the balance test using balance test scores constructed from internal data only (with cross-fitting) and from external data under Setting~2 (moderate distribution shift). The internal and external sample sizes are equal ($N=500$).}
\label{tab: external balance power comparison}
\resizebox{\textwidth}{!}{%
\begin{tabular}{cccccccc}
\toprule
& & \multicolumn{2}{c}{Model 1} & \multicolumn{2}{c}{Model 2} & \multicolumn{2}{c}{Model 3} \\
\cmidrule(rl){3-4} \cmidrule(rl){5-6} \cmidrule(rl){7-8}
 & Data Setting & Scenario 1 & Scenario 2 & Scenario 1 & Scenario 2 & Scenario 1 & Scenario 2 \\
\midrule
\multirow{2}{*}{$\Gamma=2$} & Internal Data Only & 0.816 & 0.763 & 1.000 & 0.998 & 0.983 & 0.966\\
& External Data (Setting 2) & 0.570 & 0.457 & 0.999 & 0.988 & 0.945 & 0.880\\
\midrule
\multirow{2}{*}{$\Gamma=3$} & Internal Data Only & 0.192 & 0.142 & 0.933 & 0.830 & 0.634 & 0.487\\
& External Data (Setting 2) & 0.034 & 0.028 & 0.864 & 0.650 & 0.399 & 0.239\\
\bottomrule
\end{tabular}
}
\end{table}

Putting these results together, external data can be a useful alternative to cross-fitting when they are genuinely comparable to the internal matched sample, because they allow the balance scores to be treated as fixed with respect to treatment assignment while avoiding fold-specific testing and the associated multiplicity adjustment. However, when external data exhibit a distribution shift, propensity score estimates transferred from the external sample can be systematically misspecified for the internal sample. This misspecification can reduce the ability of the balance test to reject understated values of $\Gamma$ and can also attenuate the power gains in the outcome test that would otherwise arise from avoiding cross-fitting. Thus, external data are most attractive when transportability is plausible, whereas cross-fitting on the internal matched sample remains a safe choice when high-quality external data are unavailable.

\section*{Appendix G: Simulation Results under Different Standardized-Difference Thresholds}\label{sec: thresholds}

In the simulation studies in Section~\ref{sec: simulations} of the main text and Appendices B--F, we proceed to outcome analysis (including sensitivity analysis) only when the matched datasets satisfy the common balance criterion that the absolute standardized difference between the treatment and control groups is below 0.2 for every measured confounder \citep{rosenbaum2020design,
pimentel2024covariate, zhu2025randomization}. Stricter thresholds, such as 0.1, have also been considered in some studies \citep{small2024protocols}. The coverage and Type-I error guarantees in Lemma~1 and Theorem~1 are established under the corresponding regularity conditions on the post-matching treatment assignment probabilities and balance and outcome test statistics; they do not automatically extend to inference conditional on passing this additional balance check based on some standardized-difference threshold. Explicitly accounting for this screening step in the theoretical analysis of either the conventional or proposed approach would require further theoretical work and is beyond the scope of this paper. In this section, we therefore examine the empirical null rejection rates and power of both approaches among matched datasets accepted under standardized-difference thresholds of 0.2 and 0.1, using the same simulation settings as in Section~\ref{sec: simulations}.

For each standardized-difference threshold and simulation setting in Section~\ref{sec: simulations}, we generate candidate datasets and apply the same matching procedure until 1,000 matched datasets satisfy the balance criterion based on the corresponding standardized-difference threshold. Table~\ref{tab: num_under} reports the number of candidate datasets required. Candidate datasets are generated using sequential random seeds starting at 1 and increasing by 1. As expected, the stricter threshold requires more candidate datasets, with substantial increases in some simulation settings. Because the two standardized-difference thresholds retain different collections of generated datasets, comparisons across standardized-difference thresholds should not be interpreted as isolating the effect of improving post-matching covariate balance within a fixed dataset.

\begin{table}[ht]
\centering
\footnotesize
\caption{Number of datasets generated to obtain 1,000 accepted datasets under absolute standardized difference (Std) thresholds of 0.1 and 0.2 for every covariate, using sequential random seeds starting at 1 and increasing by 1.}
\resizebox{\textwidth}{!}{\begin{tabular}{cccccccc}
\toprule
& \multirow{2}{*}{}&\multicolumn{2}{c}{Model 1} &\multicolumn{2}{c}{Model 2} &\multicolumn{2}{c}{Model 3} \\ 
\cmidrule(rl){3-4} \cmidrule(rl){5-6} \cmidrule(rl){7-8}
 & Std Threshold & Scenario 1 & Scenario 2 &  Scenario 1 & Scenario 2 & Scenario 1 & Scenario 2 \\
\midrule
\multirow{2}{*}{Sample Size $N=500$} & 0.2 & 1174 & 1136 & 1881 & 1585 & 1341 & 1268 \\
 & 0.1 & 10372 & 8492 & 74755 & 51085 & 25861 & 19452 \\
\midrule
\multirow{2}{*}{Sample Size $N=1000$} & 0.2 & 1008 & 1004 & 1120 & 1086 & 1038 & 1016 \\
 & 0.1 & 3969 & 3357 & 44859 & 22758 & 11418 & 7526 \\
\bottomrule
\end{tabular}}
\label{tab: num_under}
\end{table}

We also evaluate the simulated Type-I error rates (i.e., rejection rates under the null effect) of the conventional and proposed approaches under standardized-difference thresholds of 0.2 and 0.1, using the simulation settings in Section~\ref{sec: simulations}. For both approaches, the rejection rates under the null effect are close to zero and below the nominal level of 0.05 across all settings considered. These results provide empirical support for Type-I error control in the settings examined and complement the subsequent power comparison in Tables~\ref{tab: more simulated power} and \ref{tab: power01}. It is worth clarifying that the near-zero rejection rates under the null effect and the simulated measured and unmeasured confounders do not necessarily imply that either the conventional or proposed approach is unnecessarily conservative over the full sensitivity model (i.e., over the full feasible set for unmeasured confounders). Such low rejection rates under the null effect can occur when the prespecified sensitivity parameter $\Gamma$ does not underestimate the true magnitude of confounding and the actual allocation of unmeasured confounders is less adverse than the worst-case allocation permitted by the Rosenbaum sensitivity bounds (\ref{eqn: Rosenbaum bounds}); see also the discussion in Section~3.5 of \citet{heng2020finding}.

We next examine the power gains of the proposed approach under standardized-difference thresholds of 0.2 and 0.1. Table~\ref{tab: power01} reports the simulated power of the conventional and proposed approaches under the threshold of 0.1, using the same simulation settings as Table~\ref{tab: more simulated power} (i.e., simulated power based on the threshold of 0.2) in Section~\ref{sec: simulations} of the main text. Together, the results in Tables~\ref{tab: more simulated power} and \ref{tab: power01} show that the proposed approach offers power gains over the conventional approach under both thresholds in the settings considered, with substantial gains in several simulation settings.

\begin{table}[H]
\centering
\footnotesize
\caption{Simulated power of the two approaches to Rosenbaum-type sensitivity analysis: the conventional approach and the proposed new approach, using matched datasets under absolute standardized difference thresholds of 0.1. }
\resizebox{\textwidth}{!}{%
\begin{tabular}{cccccccc}
\toprule
& \multirow{2}{*}{Sample Size $N=500$}&\multicolumn{2}{c}{Model 1} &\multicolumn{2}{c}{Model 2} &\multicolumn{2}{c}{Model 3} \\ 
\cmidrule(rl){3-4} \cmidrule(rl){5-6} \cmidrule(rl){7-8}
 & & Scenario 1 & Scenario 2 &  Scenario 1 & Scenario 2 & Scenario 1 & Scenario 2 \\
\midrule
\multirow{3}{*}{$\Gamma=7$} & Conventional & 0.352 & 0.590 & 0.449 & 0.788 & 0.404 & 0.714\\
& Proposed (XGBoost) & 0.402 & 0.649 & 0.742 & 0.915 & 0.544 & 0.797\\
& Proposed (Random Forest) & 0.426 & 0.668 & 0.736 & 0.933 & 0.560 & 0.806\\
\midrule
\multirow{3}{*}{$\Gamma=8$} & Conventional & 0.154 & 0.341 & 0.216 & 0.550 & 0.174 & 0.430\\
& Proposed (XGBoost) & 0.181 & 0.383 & 0.450 & 0.732 & 0.268 & 0.525\\
& Proposed (Random Forest) & 0.186 & 0.396 & 0.462 & 0.742 & 0.280 & 0.539\\
\midrule
& \multirow{2}{*}{Sample Size $N=1000$}&\multicolumn{2}{c}{Model 1} &\multicolumn{2}{c}{Model 2} &\multicolumn{2}{c}{Model 3} \\ 
\cmidrule(rl){3-4} \cmidrule(rl){5-6} \cmidrule(rl){7-8}
 & & Scenario 1 & Scenario 2 &  Scenario 1 & Scenario 2 & Scenario 1 & Scenario 2 \\
\midrule
\multirow{3}{*}{$\Gamma=9$} & Conventional & 0.548 & 0.835 & 0.678 & 0.973 & 0.633 & 0.921\\
& Proposed (XGBoost) & 0.690 & 0.897 & 0.977 & 0.999 & 0.855 & 0.972\\
& Proposed (Random Forest) & 0.755 & 0.930 & 0.990 & 0.999 & 0.896 & 0.988\\
\midrule
\multirow{3}{*}{$\Gamma=10$} & Conventional & 0.303 & 0.621 & 0.436 & 0.891 & 0.377 & 0.779\\
& Proposed (XGBoost) & 0.417 & 0.706 & 0.893 & 0.992 & 0.645 & 0.899\\
& Proposed (Random Forest) & 0.485 & 0.778 & 0.913 & 0.994 & 0.700 & 0.931\\
\midrule
\multirow{3}{*}{$\Gamma=11$} & Conventional & 0.133 & 0.364 & 0.241 & 0.728 & 0.182 & 0.562\\
& Proposed (XGBoost) & 0.198 & 0.458 & 0.690 & 0.945 & 0.410 & 0.743\\
& Proposed (Random Forest) & 0.248 & 0.527 & 0.743 & 0.960 & 0.469 & 0.797\\
\bottomrule
\end{tabular}}
\label{tab: power01}
\end{table}

\section*{Appendix H: Summary}\label{sec: summary}

The Rosenbaum-type sensitivity analysis (\citealp{Rosenbaum1987, rosenbaum2002observational}) is one of the most influential and widely used sensitivity analysis frameworks in causal inference, especially in matched observational studies. It quantifies the impact of post-matching confounding bias (i.e., a combined effect of residual overt bias and hidden bias after matching) on causal conclusions. The popularity of the Rosenbaum-type sensitivity analysis is largely due to its two remarkable merits: First, its statistical validity does not require any modeling assumptions on propensity scores. Second, it only involves a single, elegant sensitivity parameter $\Gamma$ (defined in the Rosenbaum bounds constraint (\ref{eqn: Rosenbaum bounds})), making the sensitivity analysis results interpretable and transparent.

In this work, we pointed out that when the matching was inexact (typically when there are continuous or multiple measured confounders), the Rosenbaum-type sensitivity analysis can be overly conservative and possibly overstate the sensitivity of causal conclusions to post-matching confounding bias. This is because the classical solution to the Rosenbaum-type sensitivity analysis (i.e., the allocation of hidden bias under the Rosenbaum bounds constraints that corresponds to the worst-case $p$-value) often contradicts the overt bias observable from the matched dataset. To address this critical limitation of the Rosenbaum-type sensitivity analysis and to make matched observational studies more robust to post-matching confounding bias, we propose a novel approach to Rosenbaum-type sensitivity analysis that removes such a contradiction between overt bias and hidden bias in matched datasets. The core idea of our approach is to use post-matching overt bias as a valid negative control to refine the feasible set of hidden bias allowed under the Rosenbaum bounds constraint. The statistical validity of our approach does \textit{not} depend on any additional modeling assumptions. We demonstrate that our proposed formulation can be efficiently solved using an iterative convex programming algorithm and can be applied to general matching designs. Through theoretical analyses, simulation studies, and a real data application, we show that our method can substantially enhance power of sensitivity analysis. Importantly, as discussed in Remark~\ref{rem: extension} in Appendix B, our approach can be extended to many other sensitivity analysis frameworks.

\putbib[references]  
\end{bibunit}

\end{document}